\documentclass[pre,floatfix,twocolumn,showpacs,superscriptaddress]{revtex4}

\usepackage{mathtools}
\usepackage{amsfonts}
\usepackage{amsmath}
\usepackage{graphicx}
\usepackage[usenames,dvipsnames]{color}
\usepackage{amssymb}
\usepackage{longtable}
\usepackage{cancel}
\usepackage{undertilde}
\usepackage{bm} 
\usepackage{bbm} 
\usepackage{tabularx}
\usepackage{makeidx,xspace,color,listings}

\allowdisplaybreaks

\newcommand{\beq}{\begin{equation}} 
\newcommand{\eeq}{\end{equation}}
\newcommand{\bqa}{\begin{eqnarray}} 
\newcommand{\eqa}{\end{eqnarray}}
\newcommand{\nn}{\nonumber}

\newcommand{\dg}{^\dagger}
\newcommand{\rt}[1]{\sqrt{#1}\,}
\newcommand{\smallfrac}[2]{\mbox{$\frac{#1}{#2}$}}
\newcommand{\half}{\smallfrac{1}{2}}
\newcommand{\bra}[1]{\langle{#1}|} 
\newcommand{\ket}[1]{|{#1}\rangle}
\newcommand{\ip}[2]{\langle{#1}|{#2}\rangle}
\newcommand{\op}[2]{|{#1}\rangle \langle{#2}|}

\newcommand{\tbt}[4]{\left(\begin{array}{cc} {#1} & {#2} \\ {#3} & {#4} \end{array}\right)}

\newcommand{\mapK}{{\cal K}}

\newcommand{\mapU}{{\cal U}}
\newcommand{\mapV}{{\cal V}}

\newcommand{\genL}{{\cal L}}
\newcommand{\genR}{{\cal R}}
\newcommand{\genS}{{\cal S}}

\newcommand{\LH}{{\cal L}_{\rm H}}
\newcommand{\LK}{{\cal L}_{\rm K}}
\newcommand{\LJH}{{\cal L}_{\rm JH}}
\newcommand{\LQW}{{\cal L}_{\rm QW}}

\newcommand{\Hhat}{\hat{H}}
\newcommand{\Uhat}{\hat{U}}

\newcommand{\Vhat}{\hat{V}}
\newcommand{\Mhat}{\hat{M}}

\newcommand{\Qs}{\hat{Q}_{\rm S}}
\newcommand{\Qt}{\hat{Q}_{\rm T}}
\newcommand{\ks}{k_{\rm S}}
\newcommand{\kt}{k_{\rm T}}

\newcommand{\Qj}{\hat{Q}_j}
\newcommand{\Qk}{\hat{Q}_k}
\newcommand{\Qjk}{\hat{Q}_{jk}}
\newcommand{\Qkj}{\hat{Q}_{kj}}

\newcommand{\Pk}{\hat{P}_k}

\newcommand{\mapW}{{\cal W}}
\newcommand{\mapM}{{\cal M}}

\newcommand{\ketg}{\ket{\psi_1}}
\newcommand{\kets}{\ket{\psi_2}}

\newcommand{\Dt}{\Delta t}

\begin{document}

\title{Coherent chemical kinetics as quantum walks I: Reaction operators for radical pairs}

\author{A. Chia}
\affiliation{Centre for Quantum Technologies, National University of Singapore}

\author{A. G\'orecka}
\affiliation{Division of Physics and Applied Physics, School of Physical and Mathematical Sciences, Nanyang Technological University, Singapore}

\author{K. C. Tan}
\affiliation{Centre for Quantum Technologies, National University of Singapore}
\affiliation{Center for Macroscopic Quantum Control, Department of Physics and Astronomy, Seoul National University}

\author{{\L}. Pawela}
\affiliation{Institute of Theoretical and Applied Informatics, Polish Academy of Sciences}

\author{P. Kurzy\'nski}
\affiliation{Centre for Quantum Technologies, National University of Singapore}
\affiliation{Faculty of Physics, Adam Mickiewicz University}

\author{T. Paterek}
\affiliation{Centre for Quantum Technologies, National University of Singapore}
\affiliation{Division of Physics and Applied Physics, School of Physical and Mathematical Sciences, Nanyang Technological University, Singapore}

\author{D. Kaszlikowski}
\affiliation{Centre for Quantum Technologies, National University of Singapore}

\date{\today}

%

\begin{abstract}

Classical chemical kinetics use rate-equation models to describe how a reaction proceeds in time. Such models are sufficient for describing state transitions in a reaction where coherences between different states do not arise, or in other words, a reaction which contain only incoherent transitions. A prominent example reaction containing coherent transitions is the radical-pair model. The kinetics of such reactions is defined by the so-called reaction operator which determines the radical-pair state as a function of intermediate transition rates. We argue that the well-known concept of quantum walks from quantum information theory is a natural and apt framework for describing multisite chemical reactions. By composing Kraus maps that act only on two sites at a time, we show how the quantum-walk formalism can be applied to derive a reaction operator for the standard avian radical-pair reaction. Our reaction operator predicts a recombination dephasing rate consistent with recent experiments [J. Chem. Phys. {\bf 139}, 234309 (2013)], in contrast to previous work by Jones and Hore [Chem. Phys. Lett. {\bf 488}, 90 (2010)]. The standard radical-pair reaction has conventionally been described by either a normalised density operator incorporating both the radical pair and reaction products, or by a trace-decreasing density operator that considers only the radical pair. We demonstrate a density operator that is both normalised and refers only to radical-pair states. Generalisations to include additional dephasing processes and an arbitrary number of sites are also discussed.

\end{abstract}


\maketitle


\section{Introduction}

It is known that many animal and insect species are capable of sensing extremely weak magnetic fields. Of particular interest amongst biologists, chemists, and physicists is the problem of how migrating species of birds use the Earth's magnetic field for navigation \cite{Rit11}. The mechanism granting birds the ability to use the geomagnetic field for guidance is known as the avian compass and there is now substantial evidence that certain species (e.g.~the European robin) do indeed possess this compass. To date, the most promising model of the avian compass is known as the radical-pair mechanism \cite{WTMH01,RTPWW04,MHC+08,TH04,RH09}, a chemical reaction which takes place inside a photoreceptor molecule in the bird's eye known as cryptochrome \cite{LMH+07,MJBL+04,MSWS04,NDG+11,LM10,WW14}. The radical-pair model has been a platform on which many interesting theoretical investigations have been carried out since it was first proposed as a candidate explanation for the avian compass \cite{SSW78,RAS00}. One interesting problem, which has also been the subject of recent debate, is the form of its reaction operator. We will use quantum walks, which is essentially an elaborate theory of Kraus maps \cite{NC10,Kra71}, to shed some light on this topic. This illustrates that quantum walks is a suitable framework for describing coherent chemical kinetics.

The description of radical-pair reactions has conventionally been that of Haberkorn's from 1976 \cite{Hab76}. This approach is phenomenological and based on arguing which of two existing differential equations for the radical-pair state should be preferred. The first is proposed by Johnson and Merrifield \cite{JM70}, and Evans {\it et al.} \cite{EFL73}. The second is due to Pedersen and Freed \cite{PF73}. We shall refer to Refs.~\cite{JM70,EFL73} collectively as the Johnson--Evans equation for mere convenience. Haberkorn chose the Johnson--Evans equation instead of the one due to Pedersen and Freed by showing that Pedersen and Freed's equation leads to negative eigenvalues for the radical-pair state whereas the Johnson and Evans version does not. Both differential equations for the radial-pair state can now be seen to contain terms making up what is known as the Lindblad form of master equations \cite{Lin76,GKS76}, though neither is actually in Lindblad form. Haberkorn's solution was to simply consider what the system state would look like if propagated using the two proposed state derivatives. The nontriviality in distinguishing the two types of state evolution lies in the fact that it is not immediately obvious how to interpret the proposed state derivatives as opposed to the state itself. In more general and formal language, Haberkorn made his argument by referring to the map ${\cal K}(t)$ which takes a system state from $\rho(0)$ to $\rho(t)$, rather than the map's generator ${\cal L}$, which is related to the map by ${\cal K}(t)=\exp({\cal L}\,t)$. Hence the generator hence defines the derivative of the state, i.e. $\dot{\rho}(t)={\cal L}\rho(t)$. In the context of chemical kinetics, the superoperator ${\cal L}$ is referred to as the reaction operator. Of course, today, the Lindblad form is well understood (at least in the quantum optics and quantum information community), so the problem pointed out by Haberkorn may not be perceived by some to be significant. However, there is still something to be learnt from Haberkorn's work, which is that when considering non-unitary evolution the map ${\cal K}(t)$ may be a more intuitive quantity to consider than the state derivative ${\cal L}$ because $\rho(t)$ is given explicitly by ${\cal K}(t)$ but only implicitly by ${\cal L}$. One could in fact argue that this is also the reason why Lindblad's result on the form of master equations is nontrivial \cite{Lin76,GKS76}. This motivates us to use $\mapK(t)$ instead of $\genL$ in this paper.

Haberkorn's preferred reaction operator went unchallenged until recently \cite{Kom09}, bringing the debate about its form into the limelight again \cite{Kom09,IPLM10,Shu10,Pur10,JH10,JMH11,TSPB12,CGTB14}. This has resulted in one side arguing in defence of Haberkorn and is now referred to as the conventional, phenomenological, or Haberkorn approach to radical-pair reactions \cite{IPLM10,Shu10,Pur10}. A separate camp, called the quantum-measurement approach to radical pairs has proposed two new reaction operators, due separately to Kominis \cite{Kom09} and Jones and Hore \cite{JH10}. Of particular interest to us is the paper by Jones and Hore \cite{JH10} who derived their reaction operator using Kraus maps \cite{Kra71}. The Jones--Hore equation predicts a different singlet-triplet dephasing rate to the conventional approach of Haberkorn's and has been the subject of a recent experiment aimed at discriminating the two models \cite{MLGH13}. This experiment showed the Jones--Hore equation to be inconsistent with the measured dephasing rate. In this paper we also use Kraus maps to describe the radical-pair kinetics but we obtain a dephasing rate that is consistent with Ref.~\cite{MLGH13}. A key factor in our approach is the recognition that any intermediate transition in a multistate reaction involve only two states at once. Although this seems trivial, it is what separates our paper from the work of Jones and Hore because it implies that one can derive the map for a multistate reaction by composing two-state maps only. Maps for multistate reactions derived in this way will thus be correct by virtue of the method (provided that we have the correct two-state maps). Two-state transitions are particularly well understood in quantum information theory since qubits, which are essentially two-state systems, are the central object of study. The toolbox provided by quantum information theory allows us to construct maps for multistate reactions that are robust to modelling errors.

Our approach to the radical-pair reaction kinetics views the reaction as simply a system evolving between a discrete set of states in a probabilistic manner. Since such systems are analogous to random walks, our approach to the problem is one of \emph{quantum} walks. We review quantum walks below and point out the sense in which our version of quantum walks differ from those in the quantum-walk literature.

Quantum walk was first introduced in the early 90s by Aharanov and coworkers \cite{ADZ93}. They sought to generalise the idea of a classical walker who can only move left or right in discrete units along one spatial dimension to the quantum case. They managed to define a quantum walk as the analogue of a classical random walk by correlating the system's spatial coordinate to its internal degree of freedom such as spin, generically called a coin. The coin's ability to be in a superposition of states can be seen to give rise to quantum walks although the use of a coin is actually not necessary \cite{PRR05}. Since then quantum walks have proven to be useful in quantum information \cite{Kem03a} where they have found a variety of algorithmic applications \cite{Amb03} in hitting \cite{CFG02,Kem03b} and searching \cite{SKW03,CG04}. Quantum walks were first introduced for closed systems which follow unitary evolution, but have recently been generalised to open systems which follow nonunitary evolution, called open quantum walks \cite{APS12,APSS12}. Such evolution is described by a map that is completely positive and trace preserving \cite{NC10}, and like their unitary counterpart, the maps defined in Refs.~\cite{APS12,APSS12} include changes in the internal degrees of freedom of the open system. To model the radical-pair reaction we propose an evolution map which makes no reference to any internal degrees of freedom. In this sense our model of the radical-pair reaction is similar to Ref.~\cite{PRR05} with the exception that we allow for nonunitary evolution.

The rest of the paper is organised as follows. In Sec.~\ref{ReactOpInLit} we define the standard radical-pair reaction and review the different reaction operators that have so far been proposed. These results form the backdrop against which our approach to radical-pair kinetics should be considered. We then introduce Kraus maps in Sec.~\ref{ReactOpFromQW} which writes maps in a particular form known as the operator-sum representation \cite{NC10}. Here maps describing processes known as amplitude-damping, dephasing, and unitary evolution are explained. We give a detailed exposition of the amplitude-damping map in Appendix~\ref{AppAD} to illustrate how the operator-sum representation provides insight to the process which would not have come by so easily if the process was described using a Lindblad-form master equation. This forms our toolbox for describing radical-pair kinetics and is used in Secs.~\ref{ReactOpStdRPM} and \ref{MoreGeneralQWs} to derive a reaction operator for the standard radical-pair reaction. In Sec.~\ref{ReactOpStdRPM} we focus on reaction operators which can be described with only amplitude-damping maps. This corresponds to the reaction operators reviewed in Sec.~\ref{ReactOpInLit} where only recombination processes are assumed to occur. An interesting result here is the radical-pair density operator obtained from a partial trace over the chemical products. This gives a new state in the radical-pair basis which has been overlooked in previous models. In Sec.~\ref{MoreGeneralQWs} we generalise the results of Sec.~\ref{ReactOpStdRPM} to include dephasing and unitary dynamics and comment on the relation between our quantum-walk approach and previous work. Finally we summarise our results in Sec.~\ref{Conclusion} and discuss its connection with a sequel paper and other relevant literature.


\section{Reaction operators in the literature}
\label{ReactOpInLit}

\subsection{The standard radical-pair reaction}
\label{StdRadicalPairReact}

\begin{figure}[t]
\centerline{\includegraphics[width=0.35\textwidth]{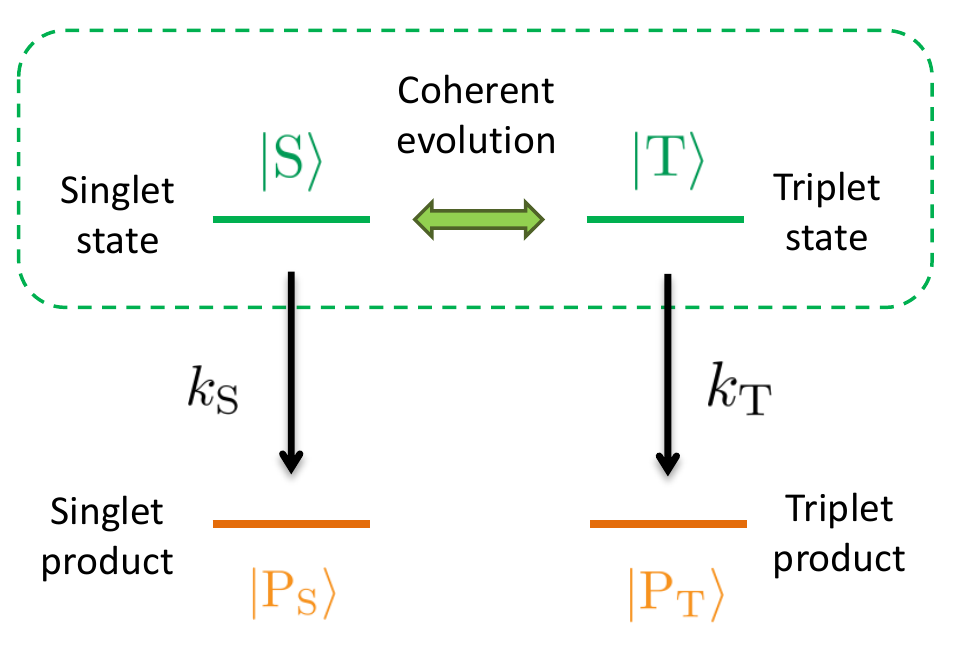}}
\caption{\label{StdRPMrates} The standard radical-pair reaction. The spin state of the radical pair oscillates coherently between singlet and triplet states under the hyperfine interaction with neighbouring nuclei spins. This oscillation can be modulated by an external magnetic field and shown to be sensitive to the magnetic field's direction. This effectively modulates the amount of time the radical pair spends in the singlet state versus the triplet. Since each spin state decays to their own unique product, information about the magnetic field direction is then encoded in the concentrations of the singlet and triplet products.}
\end{figure}

The body of literature discussed here refers to the standard radical-pair reaction shown in Fig.~\ref{StdRPMrates}. This is often referred to as a spin-selective recombination reaction \cite{RH09}. It is spin selective because the reaction product depends on the spin state of the reactants, i.e.~the radical pair, while the term recombination refers to the physical process by which the chemical products are obtained: The radical pair is typically formed through the transfer of an electron from one molecule, called the donor, to another molecule, the acceptor, creating a spatially separated pair of entangled spins. The formation of the chemical products usually involve a back-transfer of this electron from the acceptor to the donor---a recombination of the electron with the vacancy left on the donor molecule (see e.g.~Refs.~\cite{RH09,Rod09,TB12} for a more complete account of the radical-pair mechanism). 

In line with previous works \cite{Hab76,JH10,JMH11} we assume a minimal basis $\{ \ket{\rm S}, \ket{\rm T} \}$ for the reaction where $\ket{\rm S}$ denotes the singlet state and $\ket{\rm T}$ the triplet state with zero total spin. Physically this corresponds to the high-field limit where the triplet states with nonzero total spin are sufficiently far away in energy so that they can be safely neglected. We label the singlet and triplet product states as $\ket{{\rm P}_{\rm S}}$ and $\ket{{\rm P}_{\rm T}}$. The $\ket{\rm S} \longrightarrow \ket{{\rm P}_{\rm S}}$ and $\ket{\rm T} \longrightarrow \ket{{\rm P}_{\rm T}}$ transitions are then characterised by the respective rates $k_{\rm S}$ and $k_{\rm T}$. We have circled the singlet and triplet states in Fig.~\ref{StdRPMrates} with a dashed line to emphasise that the system comprises of only the $\ket{\rm S}$ and $\ket{\rm T}$ states. All known reaction operators are expressed in terms of these states so it will be convenient to define
\begin{align}
\label{QsAndQt}
	\Qs = \op{\rm S}{\rm S} \;, \quad  \Qt = \op{\rm T}{\rm T}  \;.
\end{align}
The green dashed boundary in Fig.~\ref{StdRPMrates} then serves to remind us that in the present discussion, $\Qs$ and $\Qt$ sums to the identity, i.e.~$\Qs+\Qt=\hat{1}$. This may be useful to keep in mind for later as our approach extends the system Hilbert space to include the product states so that $\ket{\rm S}$ and $\ket{\rm T}$ no longer form a complete set.

The problem is to determine the appropriate reaction operator which describes the changes brought upon $\rho(t)$ due solely to the spin-selective recombination taking the spin states to product states. In particular, this recombination contributes to the singlet-triplet dephasing and we would like to determine what exactly this contribution is. All other effects such as spin relaxation or effects of molecular motion are ignored. Spin-dependent interactions such as the Zeeman, hyperfine, exchange, or dipolar are also ignored. Including additional processes other than the spin-selective recombination will tend to increase the dephasing rate. This will be assumed in all of the reaction operators which are reviewed next, which all describe Fig.~\ref{StdRPMrates} but without the coherent evolution.

\subsection{Proposed reaction operators of the standard radical-pair reaction}
\label{ReactOpLitReview}

\subsubsection{Theoretical models}
\label{ReactOpModels}

Beginning with Haberkorn, the reaction operator preferred by his positivity-preserving argument for the radical-pair state is given by 
\begin{align}
\label{LH}
	\dot{\rho}(t) = {}& \LH \, \rho(t)  \nn \\[0.2cm]
                  \equiv {}& - \frac{1}{2} \; \ks  \Big[ \Qs \, \rho(t) + \rho(t) \, \Qs \Big] \nn \\ 
	                         & - \frac{1}{2} \; \kt  \Big[ \Qt \, \rho(t) + \rho(t) \, \Qt \Big]  \;.
\end{align}
Note that this is a non-trace-preserving equation for $\rho(t)$, the effect of which is to describe leakage of the singlet and triplet populations over time at the rates $\ks$ and $\kt$ respectively. Equation \eqref{LH} also predicts a damping of the singlet-triplet coherence at the rate of $(\ks+\kt)/2$. This can be seen by calculating the time derivative of the off-diagonal elements of $\rho(t)$:
\begin{align}
\label{DephasingRate}
	\dot{\rho}_{\rm ST}(t) = - \frac{1}{2} \big( \ks + \kt \big) \rho_{\rm ST}(t)  \;,
\end{align}
where $\rho_{\rm ST}(t)=\bra{\rm S}\rho(t)\ket{\rm T}$. This reaction operator was first questioned by Kominis \cite{Kom09} who argued that the radical-pair reaction is analogous to two coupled quantum dots under continuous measurement by a point contact. Kominis then derived a reaction operator using similar methods as Refs.~\cite{SM99,WUS+01,GMWS01} which gives
\begin{align}
\label{LK}
	\dot{\rho}(t) = {}& \LK \, \rho(t)  \nn \\
                  \equiv {}& \LH \, \rho(t) + \ks \, \Qs \, \rho(t) \, \Qs  + \kt \, \Qt \, \rho(t) \, \Qt   \;.
\end{align}
We have written $\LK$ in terms of $\LH$ to emphasise that the difference between \eqref{LH} and \eqref{LK} lies in the terms $\Qs\,\rho(t)\,\Qs$ and $\Qt\,\rho(t)\,\Qt$. The addition of these terms puts Kominis's result in Linblad form making the evolution trace preserving. This means that $\LK$ does not describe the loss of singlet or triplet populations as in \eqref{LH}. Kominis therefore augments his description of the radical-pair kinetics by an additional equation in which the radical-pair population is given by 
\begin{align}
\label{STPop}
	N(t+dt) = N(t) \big[ 1 - p_{\rm S}(t) - p_{\rm T}(t) \big]  \;,
\end{align}
where 
\begin{align}
	p_{\rm S}(t) = \ks {\rm Tr}\big[ \Qs \, \rho(t) \big] dt \;, \quad  p_{\rm T} = \kt {\rm Tr}\big[ \Qt \, \rho(t) \big] dt 
\end{align}
are the respective probabilities of a transition from either $\ket{\rm S}$ to $\ket{{\rm P}_{\rm S}}$, or $\ket{\rm T}$ to $\ket{{\rm P}_{\rm T}}$ in the infinitesimal interval $dt$. Equation \eqref{LK} does however predict the same dephasing rate as \eqref{LH}. This is obvious from \eqref{LK} since $\Qs\,\rho(t)\,\Qs$ and $\Qt\,\rho(t)\,\Qt$ do not contribute anything to $\bra{\rm S} \dot{\rho}(t) \ket{\rm T}$ due to the orthogonality of $\ket{\rm S}$ and $\ket{\rm T}$.

Spurred on by the measurement analogy made by Kominis, Jones and Hore \cite{JH10} attempted a derivation of the radical-pair reaction operator using the operator-sum representation (see Sec.~\ref{ReactOpFromQW}). Their result is given by
\begin{align}
\label{LJH}
	\dot{\rho}(t) = {}& \LJH \, \rho(t)  \nn \\[0.2cm]
                \equiv{}& - ( \ks + \kt ) \rho(t) \nn \\ 
                        & + \ks \, \Qt \, \rho(t) \, \Qt  + \kt \, \Qs \, \rho(t) \, \Qs  \;.
\end{align}
Similar to the phenomenological approach given by \eqref{LH}, this equation also does not preserve the trace of $\rho(t)$. As already mentioned, this is attributed to the loss of the singlet and triplet populations to the reaction products. It can be seen that \eqref{LJH} gives this loss rate at $\ks$ and $\kt$ for the singlet and triplet states  respectively as expected. However, the Jones--Hore reaction operator $\LJH$ gives a dephasing rate which is the sum of the recombination rates, i.e.~$\ks+\kt$, rather than the average predicted by $\LH$ (or $\LK$). Despite being perceived by Jones and Hore to be too small a difference to be detectable in an experiment \cite{JH10}, Maeda and colleagues have recently managed to place the two models under experimental scrutiny \cite{MLGH13} using pulsed electron paramagnetic resonance spectroscopy \cite{Sch91}. We briefly review some key elements of this experiment below.

\subsubsection{Experimental discrimination}
\label{MaedaExpt}

The experiment reported in Ref.~\cite{MLGH13} uses radical pairs in a modified version of the carotenoid-porphyrin-fullerene triad molecule of Ref.~\cite{MHC+08}. This system has previously been demonstrated to exhibit the anisotropic chemical response to Earth-strength magnetic fields required for the avian compass \cite{MHC+08,MWS+11}. This model system also minimises the singlet-triplet dephasing due to processes other than recombination and has a $\kt$ much smaller than $\ks$ for some temperatures (between 200--240\,K). This means the dephasing rate in this temperature regime is approximately $\ks/2$ according to $\LH$, and $\ks$ according to $\LJH$. The rate $\ks$ was then measured for this temperature range. It was shown that for certain temperatures (a range of about 30\,K), the dephasing rate from the Jones--Hore model lied above the measured values while the rate from Haberkorn's model always remained below. In practice there will be other uncontrollable processes which tend to increase the dephasing rate so the measured values of dephasing will not come solely from the $\ket{\rm S} \longrightarrow \ket{{\rm P}_{\rm S}}$ recombination. This means that any reaction operator must produce a dephasing rate which lies below the measured values of $\ks$ for all temperatures and therefore suggests the recombination kinetics according to $\LJH$ to be incorrect. In the next section we show how the idea of quantum walks can be used to derive a reaction operator with a dephasing rate consistent with the experimental data of Ref.~\cite{MLGH13}.


\section{Operator-sum representation}
\label{ReactOpFromQW}

Kraus published his theory of general state changes in quantum mechanics \cite{Kra71} in which he asked what form must a superoperator ${\cal G}$ take if it is to map a physically valid state $\rho$ to another physically valid state $\rho'$. Note that we have not actually mentioned time so $\rho'$ can be the state of a quantum system after some abstract operation, not necessarily a state at a particular instant in the future of $\rho$ (although we will use it to propagate the system in time). The answer to the question just posed is that ${\cal G}\rho$ must be of the Kraus form, also known as the operator-sum representation of ${\cal G}\rho$. Kraus's result therefore has the power to describe a large class of state transitions without referring to the underlying physics which makes the theory operational. This is what gives the Kraus formalism its versatility. To describe time evolution we simply associate $\rho$ with the system state at some arbitrary time $t$, and $\rho'$ at some later time $t+\Delta t$. The operator-sum representation can then be stated as
\begin{align}
\label{Kdecomp}
	\rho(t+\Dt) = {}& {\cal G}(\Delta t) \rho(t) \nn \\
	            = {}& \sum_{n=1}^{N} \hat{K}^{(n)}(\Delta t) \, \rho(t) \, \big[\hat{K}^{(n)}(\Delta t)\big]\dg  \;,
\end{align}
where the set of Kraus operators $\{ \hat{K}^{(n)}(\Delta t) \}_n$ satisfy the condition
\begin{align}
\label{KrausOperators}
	\sum_{n=1}^N \: \big[\hat{K}^{(n)}(\Dt)\big]\dg \, \hat{K}^{(n)}(\Dt) = \hat{1}  \;.
\end{align}
The theory also gives a prescription for calculating the probability that event $n$ is observed, given by
\begin{align}
\label{KrausProb}
	\wp_n(\Delta t) = {\rm Tr}\Big\{ \big[\hat{K}^{(n)}(\Dt)\big]\dg \, \hat{K}^{(n)}(\Dt) \, \rho(t) \Big\}  \;.
\end{align}
The sum in \eqref{Kdecomp} can be understood to be over states conditioned on events (indexed by $n$) that may be observed in an interval $\Delta t$, and hence its connection to measurements. Condition \eqref{KrausOperators} is then equivalent to the conservation of probability expressed by
\begin{align}
	\sum_{n=1}^N \; \wp_n(\Delta t) = 1  \;.
\end{align}
Note that \eqref{KrausProb} is simply the norm of the post-measurement state so that \eqref{Kdecomp} may also be written as an average over normalised conditioned states:
\begin{align}
	\rho(t+\Dt) = \sum_{n=1}^N \; \wp^{(n)}(\Dt) \, \rho^{(n)}(t+\Dt)  \;,
\end{align}
where
\begin{align}
\label{NormConState}
	\rho^{(n)}(t+\Dt) = \hat{K}^{(n)}(\Delta t) \, \rho(t) \, \big[\hat{K}^{(n)}(\Delta t)\big]\dg / \, \wp^{(n)}(\Dt)   \;.
\end{align}

\subsection{Amplitude damping}
\label{AmpDampMapMainText}

The first idea that we will borrow from quantum information theory to describe the radical-pair reaction is the amplitude-damping Kraus map. A derivation of this map can be found in Ref.~\cite{NC10} but is given in terms of a photon incident on a beam splitter. This is an example of a process that can be described by the amplitude-damping map but we believe an explanation involving only ideas from probability theory and basic quantum mechanics to be more direct and fitting for developing the reaction operator. We have thus included such a detailed exposition in Appendix~\ref{AppAD} and provide only a sketch of the amplitude-damping map below.

\begin{figure}
\center
\includegraphics[width=7cm]{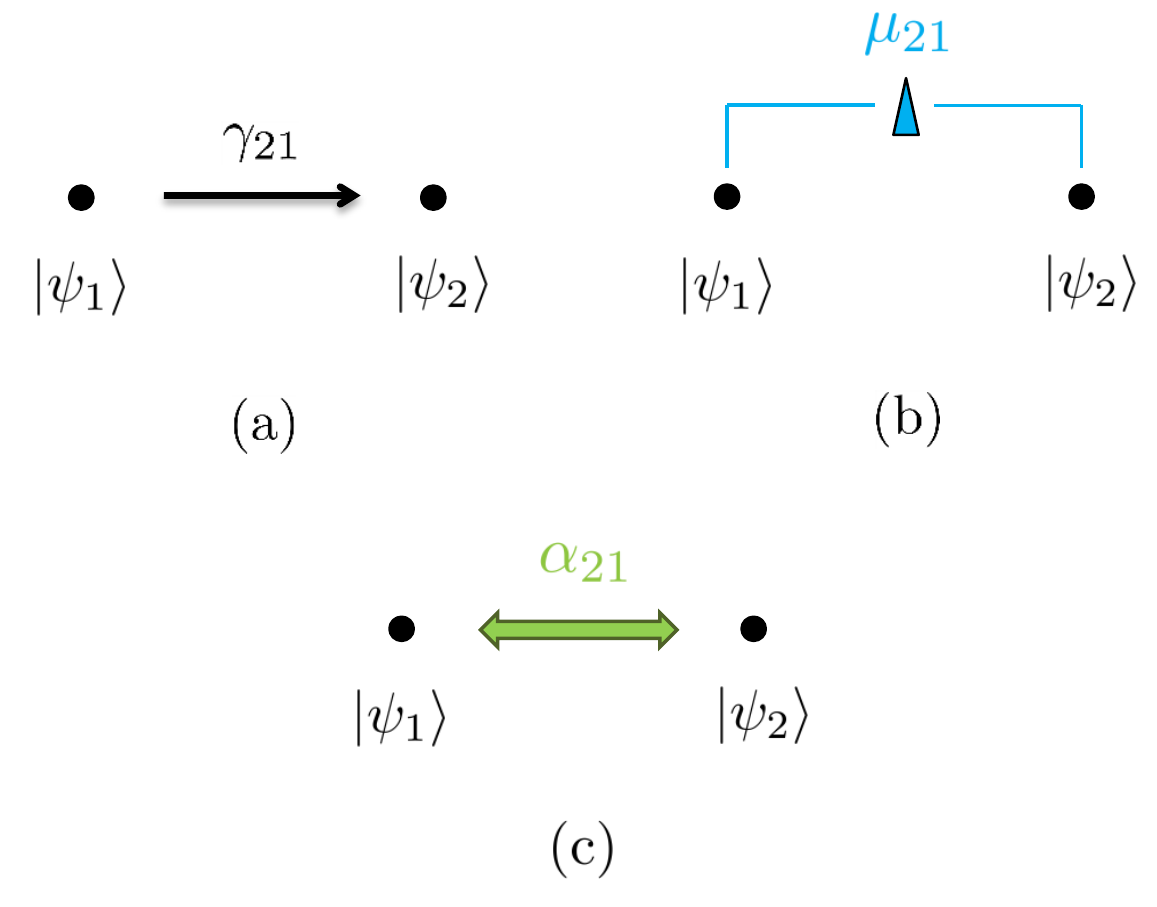} 
\caption{(a) Amplitude damping from $\ketg$ to $\kets$. This map transfers population from one state to another. (b) Dephasing between $\ketg$ and $\kets$. The map tends to localize the random walker onto $\ketg$ or $\kets$ (i.e.~turn a superposition of $\ketg$ and $\kets$ into a mixture). The triangle can be thought of as a wedge driven into a line that connects the two states being dephased. (c) Coherent oscillations between states $\ketg$ and $\kets$. The interconversion rate between states $\ketg$ and $\kets$ is given by $2\,\zeta_{21}$ [see \eqref{ParameterDefnB}].}
\label{BasicProcesses}
\end{figure}

Assuming first for simplicity that we have a two-dimensional system described by $\ket{\psi_1}$ and $\ket{\psi_2}$. Aside from its Hilbert space dimension, the system is otherwise general, and the states $\ket{\psi_1}$ and $\ket{\psi_2}$ are arbitrary. We now wish to describe the change in the system state over some time interval $\Delta t$, say from $t$ to $t+\Delta t$, allowing for the possibility that a transition from $\ket{\psi_1}$ to $\ket{\psi_2}$ may occur during this interval. This is schematically shown in Fig.~\ref{BasicProcesses}~(a). 

We can think of Fig.~\ref{BasicProcesses}~(a) as describing a possible change of state for a single molecule out of an ensemble of $n$ identical molecules. Thus some fraction of the molecules will jump from state $\ket{\psi_1}$ to $\ket{\psi_2}$ over some finite time interval. By regarding the transition as a random event we can characterise the process by a probability, $\gamma_{21}(\Delta t)$, for a given molecule to go from state $\ketg$ to $\kets$ in time $\Delta t$. If we prepare all $n$ molecules in state $\ketg$ and observe them for a time of $\Delta t$, then the fraction of molecules that make the transition to state $\kets$ is given by $\gamma_{21}(\Delta t)$. We should then expect that the longer we observe the process the greater the number of molecules to jump to state $\kets$. In the long-time limit, all $n$ molecules end up in state $\kets$ so we expect $\gamma_{21}(\Delta t) \longrightarrow 1$ as $\Delta t \longrightarrow \infty$. Conversely if the process is only observed for a very short interval then we would not expect many molecules to have jumped to state $\kets$. We thus expect $\gamma_{21}(\Delta t) \longrightarrow 0$ for $\Delta t \longrightarrow 0$. We denote the map describing such a process by $\mapM_{21}(\Dt)$. It has the operator-sum representation with two Kraus operators:
\begin{align}
\label{AmpDampMap}
	\rho(t+\Dt) = {}& \mapM_{21}(\Dt) \, \rho(t)  \nn \\[0.2cm]
	            = {}& \hat{M}^{(1)}_{21}(\Dt) \, \rho(t) \big[ \hat{M}^{(1)}_{21}(\Dt) \big]\dg  \nn \\
	                & + \hat{M}^{(2)}_{21}(\Dt) \, \rho(t) \big[ \hat{M}^{(2)}_{21}(\Dt) \big]\dg  \;,
\end{align}
where
\begin{gather}
\label{M1Dt}
	\hat{M}^{(1)}_{21}(\Dt) = \rt{\gamma_{21}(\Dt)} \, \op{\psi_2}{\psi_1}  \;,  \\
\label{M2Dt}
	\hat{M}^{(2)}_{21}(\Dt) = \op{\psi_2}{\psi_2} + \rt{1-\gamma_{21}(\Dt)} \, \op{\psi_1}{\psi_1}  \;.
\end{gather}
Its effect on an arbitrary state can be seen most directly by calculating the matrix representation of \eqref{AmpDampMap} in the basis $\{\ketg,\kets\}$. This gives the $2 \times 2$ matrix 
\begin{align}
\label{2StateAD}
	{}& \rho(t+\Delta t)  \nn \\
	  & = \tbt{\rho_{11}(t)-\gamma_{21}(\Delta t)\,\rho_{11}(t)}{\rt{1-\gamma_{21}(\Delta t)}\rho_{12}(t)}
	      {\rt{1-\gamma_{21}(\Delta t)}\rho_{21}(t)}{\rho_{22}(t) + \gamma_{21}(\Delta t)\,\rho_{11}(t)} \;,
\end{align}
where we have defined $\rho_{jk} = \bra{\psi_j} \rho \ket{\psi_k}$. The population transfer from state $\ketg$ to $\kets$ is apparent on the diagonal terms \footnote{The diagonal elements of $\rho(t)$ represent the occupation probabilities to be in each of the basis states. The actual number of molecules occupying state $\ket{\psi_k}$ at time $t$ is given by $n_k(t) = n \rho_{kk}(t)$. Assuming the total number of molecules to be conserved, $n_k(t)$ and $\rho_{kk}(t)$ differ only by a factor of $n$.} in \eqref{2StateAD}: A fraction $\gamma_{21}(\Delta t)$ has been subtracted from $\rho_{11}(t)$ and added to $\rho_{22}(t)$. Note that the off-diagonal terms of $\rho(t)$ are also affected by this process. Unless $\gamma_{21}(\Delta t)=0$ the population transfer will simultaneously reduce the coherence between $\ketg$ and $\kets$. This can be seen from Appendix~\ref{AppAD} where we argued about the form of \eqref{M1Dt} and \eqref{M2Dt} without ever referring to the system coherences. The decay of the off-diagonal elements in \eqref{2StateAD} should thus be taken as a consequence of the population transfer. This is a crucial difference between our formulation of the reaction operator and that of Ref.~\cite{JH10} where the decay of coherences was put into the system evolution by hand. When $\ketg$ represents a state of higher energy than $\kets$, \eqref{2StateAD} is said to describe a dissipative process (hence the name amplitude damping). In this case \eqref{2StateAD} captures the well-known idea from open-systems theory that dissipation implies decoherence \cite{Sch07}.

We now generalise the amplitude-damping map to a system with $N$ states. By essentially the same argument as in Appendix~\ref{AppAD}, the map describing a transition from state $\ket{\psi_k}$ to $\ket{\psi_j}$ is simply given by
\begin{align}
\label{Mjk}
	\mapM_{jk}(\Dt) \, \rho(t) = {}& \Mhat^{(1)}_{jk}(\Dt) \rho(t) \big[\Mhat^{(1)}_{jk}(\Dt)\big]\dg  \nn \\
	                               & + \Mhat^{(2)}_{jk}(\Dt) \, \rho(t) \, \big[ \Mhat^{(2)}_{jk}(\Dt) \big]\dg  \;,
\end{align}
with the Kraus operators
\begin{gather}
\label{M1General}
	\Mhat^{(1)}_{jk}(\Dt) = \sqrt{\gamma_{jk}(\Dt)} \: \Qjk \;, \\
\label{M2General}
	\Mhat^{(2)}_{jk}(\Dt) = \Pk + \sqrt{1-\gamma_{jk}(\Dt)} \: \Qk \;,  
\end{gather}
where $\gamma_{jk}(\Dt) \in [0,1]$ for all $j$, $k$, and $\Dt$. For ease of writing we have defined
\begin{gather}
	\Qjk = \op{\psi_j}{\psi_k}  \;,  \\
	\Qk = \op{\psi_k}{\psi_k} \;,  \quad  \Pk = \hat{1} - \Qk   \;.  
\end{gather}
Since $\{ \ket{\psi_k} \}_{k=1}^{N}$ is assumed to form a complete set for an arbitrary $N$, the identity operator may be written as
\begin{align}
\label{IdRes}
	\hat{1}_N = \sum_{k=1}^{N} \; \Qk  \;.
\end{align}
Note the order of subscripts in \eqref{Mjk}--\eqref{M2General} is important. Reversing the order of the subscripts reverses the direction of the process. It can be verified that \eqref{2StateAD} is reproduced by taking $j=2$, $k=1$, and $N=2$ in \eqref{Mjk}--\eqref{IdRes}. Although we have parameterised the amplitude-damping map $\mapM_{ij}(\Dt)$ by the probability $\gamma_{ij}(\Dt)$, it is usually the rate $k_{ij}$ at which the transition occurs that is measured or estimated. It is also a more useful quantity to use when expressing the amplitude-damping map in differential form. If we let $k_{ij}$ be the fraction of particles that jump from $\ket{\psi_j}$ to $\ket{\psi_i}$ in one second, then the fraction of particles that make the transition in time $\Dt$ is simply given by
\begin{align}
\label{ADRate}
	\gamma_{ij}(\Dt) = k_{ij} \; \Delta t  \;.
\end{align}

It will be useful to express \eqref{Mjk} in differential form. The two-dimensional case studied in Appendix~\ref{AppAD} can be generalised to the case of $N$ states [see \eqref{dp21/dt}]. The differential form of $\mapM_{ij}(t)$ is given by
\begin{align}
\label{AmpDampGenerator}
	\frac{d\rho}{dt} = {}& {\cal L}_{ij} \, \rho(t)  \nn \\ 
	            \equiv {}& k_{ij} \bigg[ \hat{Q}_{ij} \, \rho(t) \, \hat{Q}\dg_{ij} - \frac{1}{2} \: \Qj \, \rho(t) - \frac{1}{2} \: \rho(t) \: \Qj  \bigg]  \;.
\end{align}
This can be written in the Lindblad form if preferred by noting that $\Qj = \hat{Q}_{ji} \hat{Q}_{ij} = \hat{Q}\dg_{ij} \hat{Q}_{ij}$. We can formally express the amplitude-damping map as
\begin{align}
\label{exp(Lt)}
	\mapM_{ij}(t) = e^{{\cal L}_{ij} t}  \;,
\end{align}
where ${\cal L}_{ij}$ is said to be the generator of $\mapM_{ij}(t)$. The reader may proceed directly to Sec.~\ref{ReactOpStdRPM} at this point if he/she wishes as the results there only require knowledge of the amplitude-damping map. We will introduce the maps for dephasing and coherent evolution next but they will not appear until Sec.~\ref{MoreGeneralQWs} and hence can be read then. Lastly, note that we have suppressed the dependence of superoperators (either maps or generators) on the transition rates (or probabilities). This is because here we are thinking of the transition rate as a fixed number, a parameter which defines the system as opposed to time, which is an independent variable. In order to keep our notation simple we will continue to suppress parameter dependencies unless otherwise stated.

\subsection{Dephasing}
\label{CohEvoDeph}

We saw in \eqref{2StateAD} that population decay in one of the states led to decoherence. However, decoherence may also occur without population decay and this is known as dephasing, or phase damping in analogy to amplitude damping (so called because it results from a loss of information about the relative phases between the different basis states $\ket{\psi_k}$). This process is represented by the symbol shown in Fig.~\ref{BasicProcesses}~(b). The system evolution under dephasing over a time of $\Delta t$ can be represented simply by a $2 \times 2$ matrix for a two-state system as
\begin{align}
\label{2StatePD}
	{}& \rho(t+\Delta t)  \nn \\
	  & = {\cal V}_{21}(\Delta t) \, \rho(t)  \nn \\
	  & = \tbt{\rho_{11}(t)}{\rt{1-\mu_{21}(\Delta t)}\rho_{12}(t)}
	          {\rt{1-\mu_{21}(\Delta t)}\rho_{21}(t)}{\rho_{22}(t)} \;,
\end{align}
where $\mu_{21}(\Delta t) \in [0,1]$. Note that only the coherences (the off-diagonal terms in $\rho$) are damped. The generalisation to a system with $N$ states can be stated simply in Kraus form as
\begin{align}
\label{Vjk}
	{\cal V}_{jk}(\Dt) \, \rho(t) = {}& \Vhat^{(1)}_{jk}(\Dt) \, \rho(t) \, \big[\Vhat^{(1)}_{jk}(\Dt)\big]\dg  \nn \\
	                                  & + \Vhat^{(2)}_{jk}(\Dt) \, \rho(t) \, \big[\Vhat^{(2)}_{jk}(\Dt) \big]\dg  \;,
\end{align}
where 
\begin{gather}
\label{V1}
	\Vhat^{(1)}_{jk}(\Dt) = \sqrt{\mu_{jk}(\Dt)} \; \Qk  \;, \\
\label{V2}
	\Vhat^{(2)}_{jk}(\Dt) = \Pk + \sqrt{1-\mu_{jk}(\Dt)} \; \Qk \;.   
\end{gather}
As with amplitude damping, we can work with the rate of dephasing rather than with $\mu_{jk}(\Dt)$. Denoting the rate of dephasing between states $\ket{\psi_j}$ and $\ket{\psi_k}$ as $q_{jk}$, we can write
\begin{align}
\label{DecParam}
	\mu_{jk}(\Dt) = q_{jk} \, \Delta t  \;.
\end{align}
The evolution under the dephasing map ${\cal V}_{jk}(t)$ can then be expressed by the differential equation
\begin{align}
\label{DephasingGenerator}
	\frac{d\rho}{dt} ={}& {\cal S}_{jk} \, \rho(t)   \nn \\
              \equiv{}& q_{jk} \bigg[ \Qk \, \rho(t) \, \Qk\dg - \frac{1}{2} \: \Qk \rho(t) - \frac{1}{2} \: \rho(t) \, \Qk \bigg] ,
\end{align}
which defines the generator for ${\cal V}_{jk}(t)$:
\begin{align}
\label{DephasingExponential}
	{\cal V}_{jk}(t) = e^{{\cal S}_{jk} t}  \;.
\end{align}

\subsection{Coherent evolution}
\label{CohEvo}

The map \eqref{Vjk} describes pure decoherence and is useful for modelling processes which counter act any coherent evolution of the system that may occur in the basis $\{\ket{\psi_k}\}_{k=1}^{N}$. The singlet-triplet interconversion in the radical-pair mechanism is one such process. We depict coherent evolution between two states graphically by using a green two-way arrow as shown in Fig.~\ref{BasicProcesses}~(c). In general, coherent oscillations between states $\ket{\psi_j}$ and $\ket{\psi_k}$ can be generated by a Hamiltonian of the form (for $j \ne k$)
\begin{align}
\label{Hhatjk}
	\Hhat_{jk} = \omega_j \, \Qj + \omega_k \, \Qk + \Omega_{jk} \big( \, \Qjk + \Qkj \, \big)  \;,
\end{align}
where $\omega_k \equiv \bra{\psi_k} \Hhat_{jk} \ket{\psi_k}$ is the expectation value of $\Hhat_{jk}$ in the state $\ket{\psi_k}$. The coupling between states $\ket{\psi_j}$ and $\ket{\psi_k}$ is denoted by $\Omega_{jk}$. Note that for $\Hhat_{jk}$ to be Hermitian $\Omega_{jk}$ must be real and symmetric with respect to its indices, i.e. 
\begin{align}
	\Omega_{jk} = \Omega_{jk}^* = \Omega_{kj}  \;. 
\end{align}
Unitary evolution can be understood as a special case of the Kraus decomposition \eqref{Kdecomp} with only one Kraus operator: 
\begin{align}
	\Uhat_{jk}(\Dt) = e^{-i \Hhat_{jk} \Delta t}  \;,
\end{align}
where we have set $\hbar=1$ for convenience. The actual evolution over a time of $\Delta t$ is then effected by the map
\begin{align}
\label{Ujk}
	{\cal U}_{jk}(\Dt) \, \rho(t) = \Uhat_{jk}(\Dt) \, \rho(t) \, \Uhat\dg_{jk}(\Dt)  \;.
\end{align}
The differential form of ${\cal U}_{jk}(t)$ has the familiar commutator form:
\begin{align}
\label{UnitaryGenerator}
	\frac{d\rho}{dt} = {\cal R}_{jk} \rho(t) \equiv -i\big[ \hat{H}_{jk} , \rho(t) \big]  \;.
\end{align}
We can also express \eqref{Ujk} in terms of ${\cal R}_{jk}$ as
\begin{align}
\label{UnitaryExponential}
	{\cal U}_{jk}(t) = e^{{\cal R}_{jk} t}  \;.
\end{align}

Just as we parameterised the amplitude-damping and dephasing maps by their probability of occurrence, we can similarly parameterise unitary evolution by 
\begin{align}
\label{PrQjQk}
	\alpha_{jk}(\Dt) \equiv \big| \bra{\psi_j} \Uhat_{jk}(\Dt) \ket{\psi_k} \big|^2   \;.
\end{align}
This is the probability of making a transition to $\ket{\psi_j}$ after time $\Dt$ assuming the system was initially in $\ket{\psi_k}$. If we wish to relate $\alpha_{jk}(\Dt)$ to the transition rate under unitary evolution then an explicit expression for \eqref{PrQjQk} is required. This can be shown to be
\begin{align}
\label{PrQjQkResult}
	\alpha_{jk}(\Dt) = \frac{\Omega^2_{jk}}{2\,\zeta^2_{jk}} \; \big[ 1 - \cos\big( 2\;\!\zeta_{jk}\;\! \Dt \big) \big]  \;,
\end{align}
where we have defined
\begin{align}
\label{ParameterDefnB}
	\zeta_{jk} = \frac{1}{2} \, \rt{(\omega_k - \omega_j)^2 + 4\Omega^2_{jk}}  \;.
\end{align}
Equation \eqref{PrQjQkResult} tells us that $2\zeta_{jk}$ can be defined as the frequency at which the system oscillates between states $\ket{\psi_j}$ and $\ket{\psi_k}$. Note that this depends on both the coupling between $\ket{\psi_j}$ and $\ket{\psi_k}$ (i.e.~$\Omega_{jk}$) as well as their separation (given by $|\omega_k-\omega_j|$). We can also see from \eqref{PrQjQk} that increasing $|\omega_j-\omega_k|$ lowers the peak of the transition probability between $\ket{\psi_j}$ and $\ket{\psi_k}$. The proof of \eqref{ParameterDefnB} [and hence \eqref{PrQjQk}] will be presented in a sequel paper where it is actually used to simulate an example quantum walk. Here it is sufficient to see how $\alpha_{jk}$ is related to the rate of the process and the effect of varying $\omega_k$ and $\Omega_{jk}$.


\section{Radical-pair recombination reaction as a quantum walk}
\label{ReactOpStdRPM}

\subsection{Reaction operator}

The quantum-walk formalism visualises state transitions in a quantum system as a network of nodes (representing states) connected by edges (representing transitions), called graphs. Such models have a wide applicability because the nodes can represent abstract degrees of freedom, such as a spatial coordinate, or in our case, the state of a molecule. We therefore begin our quantum-walk model of the reaction outlined in Fig.~\ref{StdRPMrates} by simply representing the different radical-pair and product states as nodes on a graph labelled according to:
\begin{gather}
\label{SitesST}
	\ket{\rm S}\equiv\ket{\psi_1} \;, \quad  \ket{\rm T}\equiv\ket{\psi_3} \;,  \\
\label{SitesP}
	\ket{{\rm P}_{\rm S}}\equiv\ket{\psi_2} \;,  \quad  \ket{{\rm P}_{\rm T}}\equiv\ket{\psi_4} \;,
\end{gather}
while the rates are taken as
\begin{align}
\label{SitesRates}
	\ks \equiv k_{21} \;, \quad  \kt \equiv k_{43}  \;.
\end{align}
\begin{figure}[t]
\centerline{\includegraphics[width=0.19\textwidth]{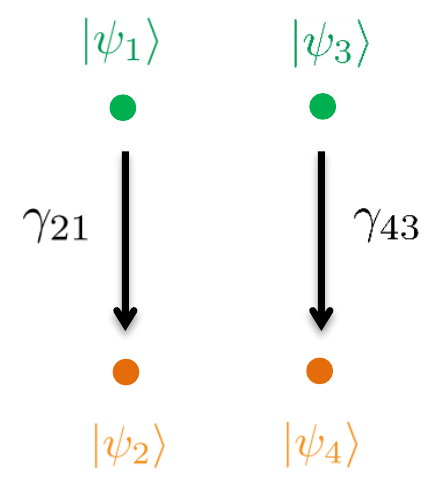}}
\caption{\label{StdRPMprob} The standard radical-pair reaction without coherent evolution represented as a graph. Black arrows represent population transfer, which here is associated with the recombination reaction of the radical pair.}
\end{figure}
The states in \eqref{SitesST} and \eqref{SitesP} are assumed to represent distinct stages of the radical-pair reaction, therefore we take $\{\ket{\psi_k}\}_{k=1}^4$ to be an orthornormal basis. This is shown in Fig.~\ref{StdRPMprob}. As explained in Sec.~\ref{ReactOpInLit}, here we concentrate only on the recombination processes which take the system from the singlet to singlet product ($\ket{\psi_1} \longrightarrow \ket{\psi_2}$), and from the triplet to the triplet product ($\ket{\psi_3} \longrightarrow \ket{\psi_4}$). Extensions to include dephasing and coherent evolution will be covered in Sec.~\ref{MoreGeneralQWs}.

Our goal is to derive a reaction operator ${\cal L}$. This means that we should express the evolution of $\rho(t)$ in differential form. As before, we can simply obtain such an equation by propagating $\rho(t)$ over an infinitesimal interval $dt$, except now there is more than one process happening at a time. This can easily be dealt with by using a single map $\mapK_{\rm QW}(dt)$ composed from a series of maps. Each map in $\mapK_{\rm QW}(dt)$ represents a particular process in the system. Attributing each recombination process to an amplitude-damping map, we describe the evolution of $\rho(t)$ by 
\begin{align}
\label{InfinitesimalEvolution}
	  \rho(t+dt) = \mapK_{\rm QW}(dt) \, \rho(t) \equiv \mapM_{43}(dt) \, \mapM_{21}(dt) \, \rho(t)  \;,
\end{align}
where on using \eqref{exp(Lt)} we have
\begin{align}	
\label{Composition}
	\mapK_{\rm QW}(dt) = e^{{\cal L}_{43} \,dt} \, e^{{\cal L}_{21} \,dt} = e^{({\cal L}_{43}+{\cal L}_{21})dt}  \;.
\end{align}
Note the second equality follows because $dt$ is infinitesimal. This can be seen by expanding the exponentials and neglecting terms on the order of $dt^2$. This also means that reversing the order of $\exp({\cal L}_{43}\;\!dt)$ and $\exp({\cal L}_{21}\;\!dt)$ does not change the second equality of \eqref{Composition}. Since ${\cal L}_{21}$ and ${\cal L}_{43}$ are independent of time, \eqref{Composition} shows that ${\cal L}_{43}+{\cal L}_{21}$ is the generator of $\mapK_{\rm QW}(t)$ for any finite $t$ [otherwise it is only the generator of $\mapK_{\rm QW}(dt)$]. This is simply equivalent to
\begin{align}
\label{ReactOpDefn}	
	\frac{d\rho}{dt} = \genL_{\rm QW} \, \rho(t) \equiv \big( {\cal L}_{21} + {\cal L}_{43} \big) \rho(t)  \;.
\end{align}
We can now apply \eqref{AmpDampGenerator} to \eqref{ReactOpDefn} to obtain
\begin{align}
\label{LQW}
	{}& \genL_{\rm QW} \, \rho(t)  \nn \\
	{}& \equiv k_{21} \bigg[ \hat{Q}_{21} \, \rho(t) \, \hat{Q}\dg_{21} 
	           - \frac{1}{2} \: \hat{Q}_{1} \, \rho(t) - \frac{1}{2} \: \rho(t) \: \hat{Q}_{1} \bigg]  \nn \\
	  & \quad + k_{43} \bigg[ \hat{Q}_{43} \, \rho(t) \, \hat{Q}\dg_{43}  
	          - \frac{1}{2} \: \hat{Q}_{3} \, \rho(t) - \frac{1}{2} \: \rho(t) \: \hat{Q}_{3} \bigg] \;.
\end{align}
A few remarks can be made about our result directly from the form of \eqref{LQW} but it will be easier to refer to its matrix representation in the site basis. This is a set of rate equations which includes the coherences between sites as well. We follow the standard convention of writing matrices where $\dot{\rho}_{mn}=\bra{\psi_m}\dot{\rho}\ket{\psi_n}$ denotes the element at the $m^{\rm th}$ row and $n^{\rm th}$ column of $\dot{\rho}$. The matrix form of \eqref{LQW} can then be easily verfied to be given by 
\begin{widetext}
\begin{align}
\label{MatrixLQW}
	\frac{d\rho}{dt} 
	= \left( \begin{array}{cccc} 
	  -k_{21} \, \rho_{11}                 &  -\half \, k_{21} \, \rho_{12}  &  -\half \, (k_{21}+k_{43}) \rho_{13}  &  -\half \, k_{21} \, \rho_{14}   \\ 
	  -\half \, k_{21} \, \rho_{21}        &  k_{21} \, \rho_{11}            &  -\half \, k_{43} \, \rho_{23}        &  0                               \\
	  -\half \, (k_{21}+k_{43}) \rho_{31}  &  -\half \, k_{43} \, \rho_{32}  &  -k_{43} \, \rho_{33}                 &  -\half \, k_{43} \, \rho_{34}   \\
	  -\half \, k_{21} \, \rho_{41}        &  0                              &  -\half \, k_{43} \, \rho_{43}        &  k_{43} \, \rho_{33}
	\end{array}\right)  \;.
\end{align}
\end{widetext}

It is obvious from the matrix form of $\LQW \rho$ that the spin-selective recombination reduces all coherences except for the coherences of the two product states, given by $\dot{\rho}_{24}$ and $\dot{\rho}_{42}$ to be zero [recall \eqref{SitesST} and \eqref{SitesP}]. Of special importance is the decay of the singlet-triplet coherence. This is given by $\dot{\rho}_{13}$ in \eqref{MatrixLQW} which can be seen to be consistent with the experiment of Ref.~\cite{MLGH13}. Note that because $\rho$ is Hermitian it follows that $\dot{\rho}$ is also Hermitian, so referring to $\dot{\rho}_{13}$ is the same as referring to $\dot{\rho}_{31}$.

Adding the diagonal elements of \eqref{MatrixLQW} we see that \eqref{LQW} is trace preserving. In particular we see that
\begin{align}
\label{ConserveLaw}
	- \frac{d\rho_{11}}{dt} = \frac{d\rho_{22}}{dt} = k_{21} \, \rho_{11}  \;,  \quad  - \frac{d\rho_{33}}{dt} = \frac{d\rho_{44}}{dt} = k_{43} \, \rho_{33}  \;.
\end{align}
That is to say, the rate at which singlet-state radical pairs are lost due to recombination is exactly balanced by the rate of increase of the singlet product. Recall from  Sec.~\ref{ReactOpLitReview} that previous treatments on the radical-pair kinetics use trace-decreasing reaction operators which refer only to $\ket{\psi_1}$ and $\ket{\psi_3}$ [see \eqref{LH} and \eqref{LJH}]. It has been noted in Ref.~\cite{JH10} that such reaction operators produce singlet and triplet populations which satisfy \eqref{ConserveLaw} and thus poses no problem. However, a description in the minimal basis $\{\ket{\psi_1}, \ket{\psi_3}\}$ still fails to account for coherences between the radical pair and product states. Jones and Hore have argued that such coherent superpositions between the radical pair and products decohere very quickly and is therefore consistent with a model in which they are neglected \footnote{See the final paragraph of Sec.~2 on page 91 of Ref.~\cite{JH10}.}. However, on accepting \eqref{MatrixLQW}, we see that coherences between the radical pair and products in fact decay at a rate less than the radical-pair dephasing (e.g.~$\dot{\rho}_{21}$ compared with $\dot{\rho}_{31}$) so the remark by Jones and Hore is not actually correct. 
Nevertheless, a model in which the product states are neglected is still permissible so long as the radical-pair populations and coherences do not depend on the populations and coherences of the products. This is clearly true from the matrix form of $\LQW$ so we can write down such a reaction operator directly. For ease of comparison with previous results we express this reaction operator in the notation of Sec.~\ref{ReactOpLitReview} [recall \eqref{QsAndQt}, \eqref{SitesST} and \eqref{SitesRates}]:
\begin{align}
\label{DefbarLQW}
	\bar{\cal L}_{\rm QW} \rho(t) 
	= {}& \dot{\rho}_{\rm SS}(t) \, \op{\rm S}{\rm S} + \dot{\rho}_{\rm ST}(t) \, \op{\rm S}{\rm T} \nn \\ 
	    & + \dot{\rho}_{\rm TS}(t) \, \op{\rm T}{\rm S} + \dot{\rho}_{\rm TT}(t) \, \op{\rm T}{\rm T}  \;.
\end{align}
We have used an overbar on $\LQW$ to indicate that it is no longer trace preserving. We can simply read off $\dot{\rho}_{\rm SS}$, $\dot{\rho}_{\rm TT}$, and $\dot{\rho}_{\rm ST}$ from \eqref{MatrixLQW} to get
\begin{align}
\label{barLQW1}
	\bar{\cal L}_{\rm QW} \rho(t)	= {}& - \ks \, \Qs \, \rho(t) \, \Qs - \frac{1}{2} \: \big(\ks + \kt\big) \, \Qs \, \rho(t) \, \Qt  \nn \\
	                                  & - \frac{1}{2} \: \big(\ks + \kt\big) \, \Qt \, \rho(t) \, \Qs - \kt \, \Qt \, \rho(t) \, \Qt  . 
\end{align}
The reader may have already noticed that $\dot{\rho}_{\rm SS}$, $\dot{\rho}_{\rm TT}$, and $\dot{\rho}_{\rm ST}$ are in fact the same as those given by the Haberkorn reaction operator $\LH$ given in \eqref{LH}, which means that \eqref{barLQW1} and \eqref{LH} should in fact be the same. This can be shown by collecting terms porportional to $\ks$ as one group and terms proportional to $\kt$ as one group in \eqref{barLQW1}:                            
\begin{align}	                               
\label{barLQW2}   
	\bar{\cal L}_{\rm QW} \rho(t) = {}& - \ks \, \Big[ \Qs \, \rho(t) \, \Qs + \frac{1}{2} \: \Qs \, \rho(t) \, \Qt  \nn \\ 
                                    & + \frac{1}{2} \: \Qt \, \rho(t) \, \Qs  \Big] - \kt \, \Big[  \Qt \, \rho(t) \, \Qt  \nn \\
                                    & + \frac{1}{2} \: \Qt \, \rho(t) \, \Qs + \frac{1}{2} \: \Qs \, \rho(t) \, \Qt \Big]  \\[0.2cm]
\label{barLQW3}   
                                = {}& - \frac{1}{2} \: \ks \Big[  \Qs \, \rho(t) + \rho(t) \, \Qs \Big]  \nn \\
                                    & - \frac{1}{2} \: \kt \Big[ \Qt \, \rho(t) + \rho(t) \, \Qt \Big]  \;.                  
\end{align}
The second equality is obtained by using $\Qt=\hat{1}_2-\Qs$ and $\Qs=\hat{1}_2-\Qt$ in the terms proportional to $\ks$ and $\kt$ respectively in \eqref{barLQW2}. Note the identity operator carries a subscript 2 because it is now only an identity on the subspace spanned by the singlet and triplet states. The resolution of the full identity operator, $\hat{1}_4$, requires all four states of the radical pair and products. This is why we have circled the singlet and triplet states in Fig.~\ref{StdRPMrates} in Sec.~\ref{StdRadicalPairReact}. We have thus derived the conventional spin-selective recombination operator using the operational and systematic treatment of quantum walks.

Writing \eqref{barLQW1} in the form of \eqref{barLQW2} also makes the comparison with the Jones--Hore reaction operator easier. This is because, as argued by Jones and Hore, an alternative derivation of their result given by \eqref{LJH} begins with
\begin{align}
\label{LJH1}
	{}& \LJH \, \rho(t) \nn \\
	{}& = - \ks \, \Big[ \Qs \, \rho(t) \, \Qs + \Qs \, \rho(t) \, \Qt  + \Qt \, \rho(t) \, \Qs  \Big]  \nn \\
	  & \quad - \kt \, \Big[  \Qt \, \rho(t) \, \Qt + \Qt \, \rho(t) \, \Qs + \Qs \, \rho(t) \, \Qt \Big]   \;.
\end{align}
The difference between \eqref{barLQW2} and \eqref{LJH1} can thus be seen in the coefficient of the cross terms [i.e.~$\Qs\,\rho(t)\,\Qt$ and $\Qt\,\rho(t)\,\Qs$]. The singlet-triplet dephasing rate was thus incorrectly posited at the outset in their argument. Alternatively, this can also be attributed to an incorrect formulation of the Kraus operators in the minimal basis. Our approach on the other hand begins with all four states of the radical pair and products. This allows us to focus on describing the $\ket{\rm S} \longrightarrow \ket{{\rm P}_{\rm S}}$ and $\ket{\rm T} \longrightarrow \ket{{\rm P}_{\rm T}}$ transitions with the dephasing between $\ket{\rm S}$ and $\ket{\rm T}$ occurring as a consequence. Since the $\ket{\rm T} \longrightarrow \ket{{\rm P}_{\rm T}}$ and $\ket{\rm S} \longrightarrow \ket{{\rm P}_{\rm S}}$ transitions are identical we in fact only have to find the correct operator-sum representation for one of them and apply it twice to $\rho(t)$ to obtain the map for the full reaction as explained in \eqref{InfinitesimalEvolution}--\eqref{ReactOpDefn}. This makes our quantum-walk approach less prone to modelling errors. Next we illustrate this point further by using the quantum-walk idea to obtain the appropriate map in the minimal basis by ignoring the chemical products.

\subsection{Ignoring chemical products}
\label{IgnoringProducts}

If we are interested in deriving the Kraus map for only the radical-pair state then we need some way of capturing the effect of the decay from the radical pair to the products but without including the products in $\rho(t)$. The above treatment of first deriving \eqref{MatrixLQW} and then reading off the equations of motion for the radical pair provides one way of achieving this result. Here we show an alternative method which is also based on quantum walks. The approach here is to find the operator-sum representation for the radical-pair state with the chemical products ignored.

\begin{figure}[t]
\centerline{\includegraphics[width=0.18\textwidth]{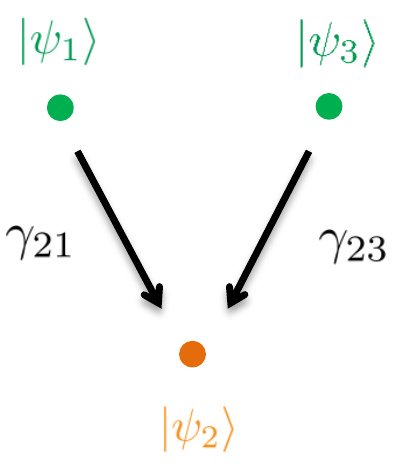}}
\caption{\label{NoChemProdGraph} A simplified graph sufficient for deriving the Kraus operator corresponding to a trace-decreasing reaction operator. Due to the decoupling of the matrix elements in \eqref{MatrixBrvLQW}, this graph can in fact be used to simulate the partial trace provided that we ignore the coherences between the product and radical pair and that we understand the distinction between $\ket{\rm P}$ and $\ket{n_1=0,n_3=0}$ [see Sec.~\ref{MainTextPartialTrace}, especially the discussion between \eqref{MatrixBrvLQW} and \eqref{NullState}].}
\end{figure}

Although it may sound contradictory, we will begin by including the product states in $\rho(t)$. However, we will lump the states $\ket{{\rm P}_{\rm S}}$ and $\ket{{\rm P}_{\rm T}}$ into a single state which we denote as $\ket{\rm P}$. This is shown in Fig.~\ref{NoChemProdGraph}. The singlet and triplet states are defined as before in \eqref{SitesST}, but instead of \eqref{SitesP} and \eqref{SitesRates} we now have
\begin{gather}
\label{Pstate}
	\ket{\rm P} \equiv \ket{\psi_2}  \;,  \\
\label{RedDephRate}
	\ks \equiv k_{21} \;, \quad  \kt \equiv k_{23}  \;.
\end{gather}
The reason for introducing Fig.~\ref{NoChemProdGraph} is twofold: First, it is easier to consider Kraus operators for a trace-preserving map. The evolution corresponding to \eqref{barLQW3} can then be extracted by using just one of the Kraus operators and hence will be trace decreasing. For this purpose it is sufficient to introduce only one additional state. This will then allow us to use the quantum-walk approach by composing two trace-preserving maps corresponding to the transitions shown in Fig.~\ref{NoChemProdGraph}. The second reason for introducing Fig.~\ref{NoChemProdGraph} has to do with what is meant by ``ignoring the products''. In the language of quantum probability, any unobserved degrees of freedom in a system can be traced over to give a so-called reduced state. This procedure is called a partial trace and is known to be a trace-preserving operation. We will find that this gives a reduced state for the radical pair that seems equivalent to Fig.~\ref{NoChemProdGraph} but is in fact subtly different. It will thus be convenient to refer to Fig.~\ref{NoChemProdGraph} for comparison. We discuss these two ideas below.

\subsubsection{Evolution from discarding products}

The operator-sum representation of \eqref{barLQW3} can be derived by considering the quantum walk shown in Fig.~\ref{NoChemProdGraph}, which can easily be described by 
\begin{align}
\label{Kpt}
	\rho(t+dt) = {\cal M}(dt) \, \rho(t) \equiv \mapM_{23}(dt) \, \mapM_{21}(dt) \, \rho(t)  \;.
\end{align}
It can be shown in general that a composition of two Kraus maps is another Kraus map. This means that we can define the Kraus operators for the total map ${\cal M}(dt)$ by substituting the operator-sum representation of $\mapM_{23}(dt)$ and $\mapM_{21}(dt)$ into \eqref{Kpt}. This gives us
\begin{align}
\label{3stateRho(t+dt)}
	{\cal M}(dt) \, \rho(t) = \sum_{n=0}^3 \; \hat{M}^{(n)}(dt) \, \rho(t) \, \big[ \hat{M}^{(n)}(dt) \big]\dg  \;,
\end{align}
where we have defined 
\begin{align}
\label{M0}
	\hat{M}^{(0)}(dt) \equiv {}& \hat{M}^{(1)}_{23}(dt) \, \hat{M}^{(1)}_{21}(dt) \;, \\
\label{M1}
	\hat{M}^{(1)}(dt) \equiv {}& \hat{M}^{(1)}_{23}(dt) \, \hat{M}^{(2)}_{21}(dt) \;, \\
\label{M2}
	\hat{M}^{(2)}(dt) \equiv {}& \hat{M}^{(2)}_{23}(dt) \, \hat{M}^{(1)}_{21}(dt) \;, \\
\label{M3}
	\hat{M}^{(3)}(dt) \equiv {}& \hat{M}^{(2)}_{23}(dt) \, \hat{M}^{(2)}_{21}(dt) \;.
\end{align}
Using \eqref{M1General} and \eqref{M2General} we find that 
\begin{align}
\label{M0dt}
	\hat{M}^{(0)}(dt) = 0  \;,
\end{align}
which means that it is redundant and we need only three Kraus operators to describe Fig.~\ref{NoChemProdGraph}. Using \eqref{ADRate} and the binomial expansion to first order in $dt$ in \eqref{M1}--\eqref{M3} [see \eqref{Binomial}] we find
\begin{gather}
\label{M1dt}
	\hat{M}^{(1)}(dt) = \rt{k_{21} \, dt} \, \hat{Q}_{21}  \;,  \\
\label{M2dt}
	\hat{M}^{(2)}(dt) = \rt{k_{23} \, dt} \, \hat{Q}_{23}  \;,  \\
\label{M3dt}
	\hat{M}^{(3)}(dt) = \hat{1} - \frac{k_{21}}{2} \: \hat{Q}_{1} \, dt - \frac{k_{23}}{2} \: \hat{Q}_{3} \, dt \;.
\end{gather}
We should note that \eqref{M1dt}--\eqref{M3dt} could have also been obtained by generalising the argument in Appendix~\ref{AppAD} to three states. In this case we would actually just write down three Kraus operators directly. It is simple to check that \eqref{M1dt}--\eqref{M3dt} satisfy the condition to be a valid set of Kraus operators [see \eqref{Kdecomp}--\eqref{NormConState} from Sec.~\ref{ReactOpFromQW}]:
\begin{align}
	\sum_{n=1}^3 \, \big[ \hat{M}^{(n)}(dt) \big]\dg \hat{M}^{(n)}(dt) = \hat{1}  \;.
\end{align}

The advantage of using \eqref{3stateRho(t+dt)} is that it decomposes $\rho(t+dt)$ into a sum of states 
\begin{align}
	\bar{\rho}^{(n)}(t+dt) \equiv \hat{M}^{(n)}(dt)\,\rho(t)\,[\hat{M}^{(n)}(dt)]\dg  \;,
\end{align}
conditioned on the outcome $n$ of a measurement performed at time $t$. The measurement is devised to give us information about the transitions occurring in the system so that we associate the $n=1$ outcome with the $\ket{\psi_1} \longrightarrow \ket{\psi_2}$ transition, $n=2$ with $\ket{\psi_3} \longrightarrow \ket{\psi_2}$, and $n=3$ with no transitions. It is easy to see that \eqref{M1dt} describes a jump from state $\ket{\psi_1}$ to $\ket{\psi_2}$ giving the conditional state
\begin{align}
\label{JumpEvo1}
	\bar{\rho}^{(1)}(t+dt) = {}& k_{21} \, \hat{Q}_{21} \, \rho(t) \, \hat{Q}\dg_{21} \, dt  \nn \\
	                       = {}& \gamma_{21}(dt) \, \rho_{11}(t) \, \op{\psi_2}{\psi_2} \;.
\end{align}
Note that $\gamma_{21}(dt) \, \rho_{11}(t)$ is the trace of $\bar{\rho}^{(1)}(t+dt)$ so that upon normalisation the evolved state is simply $\ket{\psi_2}$. Similarly applying \eqref{M2dt} we get
\begin{align}
\label{JumpEvo2}
	\bar{\rho}^{(2)}(t+dt) = {}& k_{23} \, \hat{Q}_{23} \, \rho(t) \, \hat{Q}\dg_{23} \, dt  \nn \\
	                             = {}& \gamma_{23}(dt) \, \rho_{33}(t) \, \op{\psi_2}{\psi_2} \;,
\end{align}
which is seen to describe a jump from state $\ket{\psi_3}$ to $\ket{\psi_2}$. This leaves \eqref{M3dt} which gives
\begin{align}
\label{NoJumpEvo}
	\bar{\rho}^{(3)}(t+dt) = {}& \rho(t) - \frac{1}{2} \: k_{21} \Big[ \hat{Q}_1 \, \rho(t) + \rho(t) \, \hat{Q}_1 \Big] dt  \nn \\
                                   & - \frac{1}{2} \: k_{23} \Big[ \hat{Q}_3 \, \rho(t) + \rho(t) \, \hat{Q}_3 \Big] dt  \;.
\end{align}
Note that this is the evolution given by Haberkorn's equation. The effect of \eqref{NoJumpEvo} can be seen directly by applying it to different initial states. Letting $\rho(t)=\op{\psi_1}{\psi_1}$ we find 
\begin{align}
	\bar{\rho}^{(3)}(t+dt) = \big( 1 - k_{21} \, dt \big) \op{\psi_1}{\psi_1}  \;.
\end{align}
where $1-k_{21}\,dt$ is the trace of $\bar{\rho}^{(3)}(t+dt)$. A similar result can also be seen by letting $\rho(t)=\op{\psi_3}{\psi_3}$. If the system is in the product state then we expect it to remain in the product state forever since there is no process to take the system out of $\ket{\psi_2}$. Indeed, setting $\rho(t)=\op{\psi_2}{\psi_2}$ we find 
\begin{align}
	\bar{\rho}^{(3)}(t+dt) =  \op{\psi_2}{\psi_2}  \;.
\end{align}
The evolution described by \eqref{NoJumpEvo} is thus conditioned on the absence of recombinations. That \eqref{M3dt} describes radical-pair evolution conditioned on no recombinations is not surprising in view of Appendix~\ref{AppAD} where it can be seen to be so by construction. The Kraus operator consistent with the conventional description of radical-pair kinetics in the minimal basis is thus given by \eqref{M3dt} and is the measurement approach that Jones and Hore sought after in Ref.~\cite{JH10}. That such an evolution equation is given by conditioning on unrecombined radical pairs has also been noted in Ref.~\cite{JMSH11}, but with the operator-sum formalism incorrectly applied as the experiment of Ref.~\cite{MLGH13} has now shown.


\subsubsection{Evolution from tracing out products}
\label{MainTextPartialTrace}

Here we would like to derive an equation of motion for the radical pair from tracing over the products in \eqref{LQW}. The partial trace is a formal procedure for obtaining a density operator with the products ignored and is quite different to simply discarding the products. Note that the partial trace is not simply the sum $\bra{\psi_2}\dot{\rho}(t)\ket{\psi_2}+\bra{\psi_4}\dot{\rho}(t)\ket{\psi_4}$, but rather an operation defined only on systems with a tensor product structure. As such, we require that the Hilbert space $\mathbb{H}$ of $\rho(t)$ in \eqref{LQW} be of the form
\begin{align} 
\label{FullHilbertSpace}
	\mathbb{H} = \mathbb{H}_{\rm R} \otimes \mathbb{H}_{\rm P}  \;,
\end{align}
where $\mathbb{H}_{\rm R}$ is the Hilbert space for the radical pair and $\mathbb{H}_{\rm P}$ is the space for the products. We can then derive the state of the radical pair with the products ignored from
\begin{align}
	\rho_{\rm R}(t) = {\rm Tr}_{\rm P} \big[ \rho(t) \big] \equiv \sum_{k} \; \bra{\chi_k} \rho(t) \ket{\chi_k}  \;,
\end{align}
where $\{\ket{\chi_k}\}_k$ is any basis of $\mathbb{H}_{\rm P}$.

For this reason, we will represent each site with a two-dimensional Hilbert space $\mathbb{H}_k$ ($k=1,2,3,4$) with basis $\{\ket{n_k=0},\ket{n_k=1}\}$ indicating the presence ($n_k=1$) or absence ($n_k=0$) of a random walker. Since $\ket{\psi_k}$ denotes a random walker at site $k$, we can write the site basis as
\begin{align}
\label{Psi1}
	\ket{\psi_1} = \ket{n_1=1} \otimes \ket{n_2=0} \otimes \ket{n_3=0} \otimes \ket{n_4=0}  \;,  \\
	\ket{\psi_2} = \ket{n_1=0} \otimes \ket{n_2=1} \otimes \ket{n_3=0} \otimes \ket{n_4=0}  \;,  \\
	\ket{\psi_3} = \ket{n_1=0} \otimes \ket{n_2=0} \otimes \ket{n_3=1} \otimes \ket{n_4=0}  \;,  \\
\label{Psi4}
	\ket{\psi_4} = \ket{n_1=0} \otimes \ket{n_2=0} \otimes \ket{n_3=0} \otimes \ket{n_4=1}  \;.
\end{align}
This method of defining sites can be viewed as defining a four-mode state in quantum optics with each mode containing at most one photon (or equivalently a four-mode fermionic state). Note that this expresses $\mathbb{H}$ in the form of \eqref{FullHilbertSpace} with 
\begin{align}
	\mathbb{H}_{\rm R} = \mathbb{H}_1 \otimes \mathbb{H}_3 \;, \quad \mathbb{H}_{\rm P} = \mathbb{H}_2 \otimes \mathbb{H}_4  \;.
\end{align}
It will be convenient to introduce a short-hand notation in which the tensor product is omitted. In general, we will write
\begin{align}
\label{Tensor}
	\ket{n_j,n_k} \equiv \ket{n_j} \otimes \ket{n_k}  \;.
\end{align}
The partial trace of \eqref{LQW} over sites 2 and 4 is then given by 
\begin{align}
	{}& \dot{\rho}_{\rm R}(t)  \nn \\
  {}& = \sum_{m=0}^1 \sum_{m'=0}^1 \; \bra{n_2=m,n_4=m'} \dot{\rho}(t) \ket{n_2=m,n_4=m'}  \\
\label{RhoR}
  {}& = \bra{n_2=0,n_4=0} \, \dot{\rho}(t) \, \ket{n_2=0,n_4=0} \nn \\
    & \quad + \bra{n_2=0,n_4=1} \, \dot{\rho}(t) \, \ket{n_2=0,n_4=1}  \nn \\
    & \quad + \bra{n_2=1,n_4=0} \, \dot{\rho}(t) \, \ket{n_2=1,n_4=0}  \;.
\end{align}
We have noted the expectation of $\dot{\rho}(t)$ with respect to the state $\ket{n_2=1,n_4=1}$ is identically zero since there can be at most one particle (or walker) in the system. The calculation of \eqref{RhoR} is a little bit involved so we leave the details to Appendix~\ref{PartialTrace}. The result is however simple to state. Noting that $\rho_{\rm R}$ now refers only to $\{\ket{n_1,n_3}\}_{n_1,n_3}$, we will simply denote a state with $n_1=0$ and $n_3=1$ as $\ket{0,1}$ and write
\begin{widetext}
\begin{align}
\label{ReducedLQW}
	\dot{\rho}_{\rm R}(t) = \widetilde{\genL}_{\rm QW} \, \rho_{\rm R}(t) 
                \equiv {}& \big[ k_{21} \, \rho_{11}(t) + k_{43} \, \rho_{33}(t) \big] \, \op{0,0}{0,0} 
                           - k_{43} \, \rho_{33}(t) \, \op{0,1}{0,1}  
	                         - k_{21} \, \rho_{11}(t) \, \op{1,0}{1,0}  \nn \\
	                       & - \frac{1}{2} \; \big( k_{21} + k_{43} \big) \, \rho_{13}(t) \, \op{1,0}{0,1} 
	                         - \frac{1}{2} \; \big( k_{21} + k_{43} \big) \, \rho_{31}(t) \, \op{0,1}{1,0}  \;.
\end{align}
As before, this equation is perhaps easier to read from its matrix representation, given by
\begin{align}
\label{MatrixRhoR}
	\dot{\rho}_{\rm R} 	
	=	\left( \begin{array}{ccc} 
	  k_{21} \, \rho_{11} + k_{23} \, \rho_{33}  &  0                                       &  0                                         \\ 
	  0                                          &  -k_{23} \, \rho_{33}                    &  -\half \, (k_{21}+k_{23}) \, \rho_{31}    \\
	  0                                          &  -\half \, (k_{21}+k_{23}) \, \rho_{13}  &  -k_{21} \, \rho_{11}
	\end{array} \right) 
	\equiv	\left( \begin{array}{ccc} 
	\bra{0,0}\dot{\rho}_{\rm R}\ket{0,0}  &  \bra{0,0}\dot{\rho}_{\rm R}\ket{0,1}  &  \bra{0,0}\dot{\rho}_{\rm R}\ket{1,0}   \\ 
	\bra{0,1}\dot{\rho}_{\rm R}\ket{0,0}  &  \bra{0,1}\dot{\rho}_{\rm R}\ket{0,1}  &  \bra{0,1}\dot{\rho}_{\rm R}\ket{1,0}   \\
	\bra{1,0}\dot{\rho}_{\rm R}\ket{0,0}  &  \bra{1,0}\dot{\rho}_{\rm R}\ket{0,1}  &  \bra{1,0}\dot{\rho}_{\rm R}\ket{1,0}  
	\end{array} \right)  \;.	
\end{align}
\end{widetext}

This gives the correct rates for dephasing and population transfer for the radical pair except now we have one additional state, $\ket{0,0}$, whose effect is to drain the populations out of $\ket{0,1}$ and $\ket{1,0}$. It would be tempting to compare the dynamics given by \eqref{MatrixRhoR} to the graph of Fig.~\ref{NoChemProdGraph} and define $\ket{\rm S}\equiv\ket{0,1}$, $\ket{\rm T}\equiv\ket{1,0}$, and $\ket{\rm P}\equiv\ket{0,0}$. However, the partial trace does not correspond to this identification, because as we have just shown above, the evolution defined by Fig.~\ref{NoChemProdGraph} is given by \eqref{Kpt} [or equivalently by \eqref{3stateRho(t+dt)}--\eqref{M3dt}] which has the matrix representation
\begin{widetext}
\begin{align}
\label{MatrixBrvLQW}
	\dot{\rho} 
	= \left( \begin{array}{ccc} 
	   k_{21} \, \rho_{11} + k_{23} \, \rho_{33}  &  -\half \, k_{23} \, \rho_{23}           &  -\half \, k_{21} \, \rho_{21}   \\ 
	  -\half \, k_{23} \, \rho_{32}               &  -k_{23} \, \rho_{33}                    &  -\half \, (k_{21}+k_{23}) \, \rho_{31}                                      \\
	  -\half \, k_{21} \, \rho_{12}               &  -\half \, (k_{21}+k_{23}) \, \rho_{13}  &  -k_{21} \, \rho_{11} 
	\end{array}\right)  
	\equiv	\left( \begin{array}{ccc} 
	\bra{\rm P}\dot{\rho}\ket{\rm P}  &  \bra{\rm P}\dot{\rho}\ket{\rm T}  &  \bra{\rm P}\dot{\rho}\ket{\rm S}   \\ 
	\bra{\rm T}\dot{\rho}\ket{\rm P}  &  \bra{\rm T}\dot{\rho}\ket{\rm T}  &  \bra{\rm T}\dot{\rho}\ket{\rm S}   \\
	\bra{\rm S}\dot{\rho}\ket{\rm P}  &  \bra{\rm S}\dot{\rho}\ket{\rm T}  &  \bra{\rm S}\dot{\rho}\ket{\rm S}  
	\end{array} \right)  \;.	
\end{align}
\end{widetext}
\begin{figure*}[t]
\centerline{\includegraphics[width=0.8\textwidth]{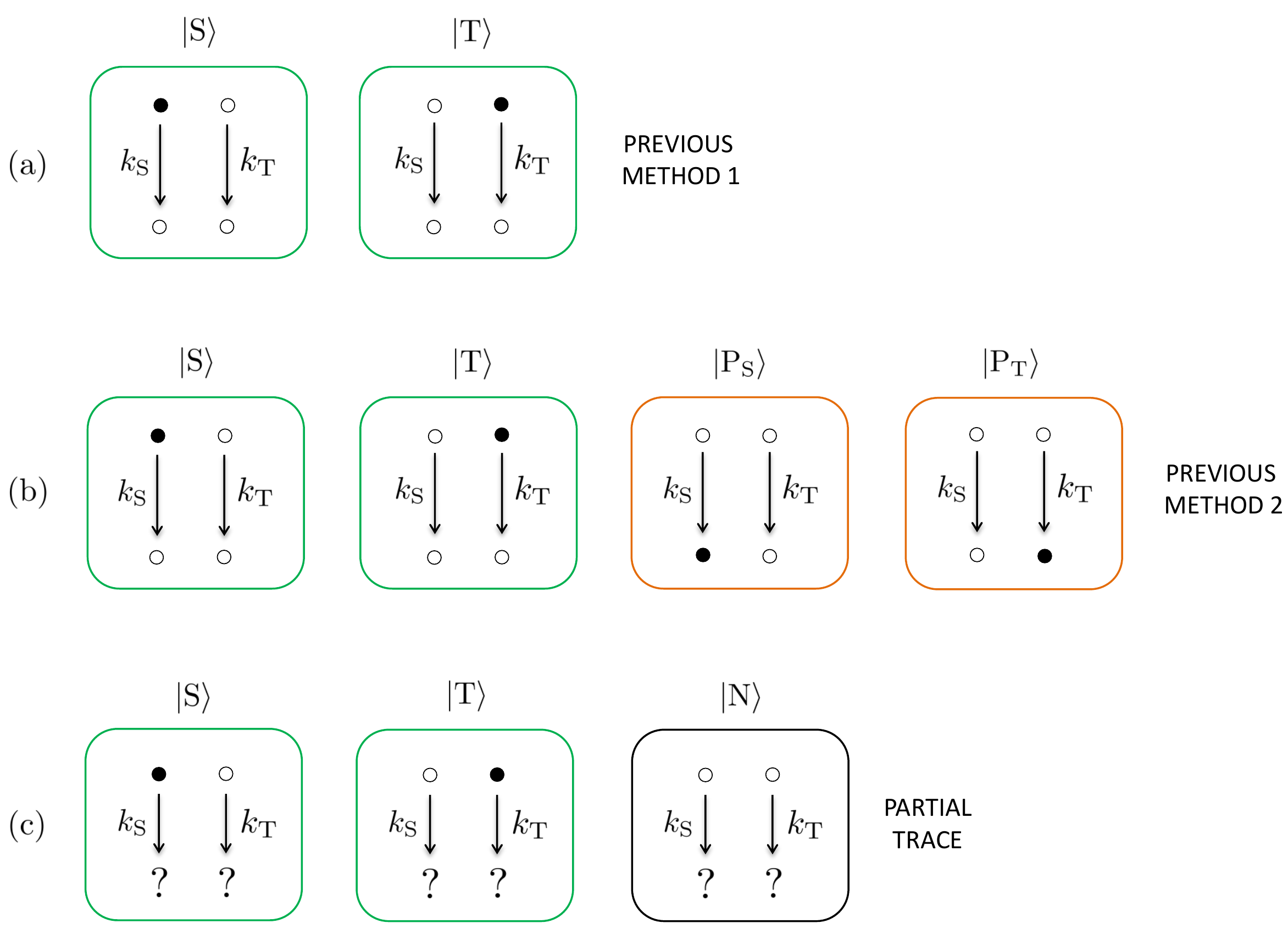}}
\caption{\label{Summary} Depiction of the different approaches to the radical-pair reaction operator based on what states are included in the model. Each box is colour coded in accordance to the colour coding of Figs.~\ref{StdRPMprob} and \ref{NoChemProdGraph}. Each site will have either a filled dot or an unfilled dot, representing respectively the presence or absence of the random walker. (a) The conventional/Haberkorn model with the reaction operator referring to only the singlet and triplet states of the radical pair. This does not preserve the norm of $\rho(t)$ and can be derived from (b) by discarding $\ket{{\rm P}_{\rm S}}$ and $\ket{{\rm P}_{\rm T}}$. (b) Including the reaction products in the reaction operator so that $\rho(t)$ is normalised at all times. (c) The partial-trace method which expresses our ignorance of the products. This gives rise to a state in which the radical pair is neither the singlet nor triplet. We represent the ignorance of products by question marks at the product sites.}
\end{figure*}
Note that we have used \eqref{SitesST} and \eqref{Pstate} and reordered the matrix elements for ease of comparison with \eqref{MatrixRhoR}. The difference between \eqref{MatrixRhoR} and \eqref{MatrixBrvLQW} is clear. Equation \eqref{MatrixBrvLQW} has coherences between the product and radical-pair states whereas \eqref{MatrixRhoR} does not. The two matrices do not even refer to the same basis---we are correct to equate $\ket{1,0}$ to $\ket{\rm S}$, and $\ket{0,1}$ to $\ket{\rm T}$ in \eqref{MatrixRhoR}, but we would be mistaken to identify $\ket{0,0}$ with $\ket{\rm P}$. The reason is because $\ket{0,0}$ conveys no other information except the absence of the random walker from sites one and three. It does not say where the walker is. Thus $\ket{0,0}$ should be regarded as a radical-pair state because it gives us only information about the radical pair, namely that it is in neither the singlet nor triplet state, consistent with the fact that we have traced over the products in \eqref{MatrixBrvLQW}. In contrast, $\ket{\rm P}$ says exactly which site the random walker is at. It is represented by a node on the graph in Fig.~\ref{NoChemProdGraph} whereas $\ket{0,0}$ is not. For notational consistency we will write 
\begin{align}
\label{NullState}
	\ket{\rm N} \equiv \ket{0,0}  \;,
\end{align}
where N may stand for ``neither'', ``none'', or ``null''. This distincition between $\ket{0,0}$ and $\ket{\rm P}$ is an important and interesting one because it suggests that $\ket{\rm N}$ is another radical-pair state that we should consider and therefore extend the minimal basis from $\{\ket{\rm S},\ket{\rm T}\}$ to $\{\ket{\rm S},\ket{\rm T},\ket{\rm N}\}$. Previous treatments on the radical-pair reaction operator have been to use either a trace-decreasing $\rho(t)$ without products, or a trace-preserving $\rho(t)$ with products. The partial trace has the advantage that it is both trace-preserving and excludes the products. It achieves this by regarding $\ket{\rm N}$ as just another radical-pair state which has not been considered (or taken seriously) before. We summarise the previous approaches to radical-pair kinetics alongside the partial-trace method schematically in Fig.~\ref{Summary}.

Finally, we make a couple of observations of the model defined by \eqref{ReducedLQW} and \eqref{MatrixRhoR}:
\begin{itemize}
\item[1.] 
We first note some similarities and differences between the partial-trace approach and the conventional Haberkorn model. The two models are essentially equivalent in that the former description uses a $3\times3$ matrix with unit trace, whereas the latter uses a $2\times2$ matrix plus one scalar equation:
\begin{align}
\label{PopSum}
	\rho_{\rm SS}(t) + \rho_{\rm TT}(t) = - \ks \int_0^t dt' \rho_{\rm SS}(t') - \kt \int_0^t dt' \rho_{\rm TT}(t')  \;.
\end{align}
We have used \eqref{ConserveLaw} to express the product populations on the right-hand side in terms of $\rho_{\rm SS}(t)$ and $\rho_{\rm TT}(t)$. However, the right-hand side of \eqref{PopSum} explicitly refers to product populations whereas the trace of $\rho_{\rm R}$ does not. This is because $\ket{\rm N}$ is a radical-pair state so that ${\rm Tr}[\rho_{\rm R}]$ is interpreted as the probability of finding the radical pair to be either a singlet, triplet, or neither. Equation \eqref{PopSum} provides more detail by saying the probability of finding the system in either the singlet, triplet, singlet product, or triplet product states must be one. Naturally, the probability of not being in the singlet or triplet state must be equal to the probability of being in the product states.
\item[2.] 
We mentioned that the random walk of Fig.~\ref{NoChemProdGraph} is not the same as the partial trace because \eqref{MatrixRhoR} and \eqref{MatrixBrvLQW} are not equivalent. However, if we note that coherences between the radical pair and products in \eqref{MatrixBrvLQW} are decoupled from the rest of the matrix, then the random walk of Fig.~\ref{NoChemProdGraph} can be used to simulate the partial trace by first defining $\ket{\psi_1}=\ket{\rm S}$, $\ket{\psi_2}=\ket{\rm N}$, and $\ket{\psi_3}=\ket{\rm T}$, and then setting $\bra{\psi_1}\rho(t)\ket{\psi_2}=\bra{\psi_3}\rho(t)\ket{\psi_2}=0$ in the end. This defines the partial trace over products in terms of a random walk. 
\end{itemize}
%


\section{Extension to more general cases}
\label{MoreGeneralQWs}

Not all processes in a chemical reaction can be represented by amplitude damping. In this section we consider reactions where coherent oscillations and additional dephasing can occur. This is the case with the radical-pair reaction where magnetic interactions give rise to coherent oscillations between the singlet and triplet states. It should be clear that \eqref{InfinitesimalEvolution}--\eqref{ReactOpDefn} can be generalised to any concatenation of maps describing amplitude damping, dephasing, or unitary evolution. More precisely, a graph defined by 
\begin{align}
\label{GeneralK}
	\mapK \equiv {}& \big( \mapU_{gh} \cdots \mapU_{cd} \, \mapU_{ab} \big)  \big( \mapV_{xy} \cdots \mapV_{rs} \, \mapV_{pq} \big)  \nn \\
	               & \times \big( \mapM_{mn} \cdots \mapM_{kl} \, \mapM_{ij} \big)  \;,
\end{align}
can be written as $\mapK(t)=\exp({\cal G} \;\!t)$ with
\begin{align}
\label{GeneralL}
	{\cal G} = {}& \big( \genR_{gh} + \cdots + \genR_{cd} + \genR_{ab} \big) + \big( \genS_{xy} + \cdots + \genS_{rs} + \genS_{pq} \big)  \nn \\
	             & + \big( \genL_{mn} + \cdots + \genL_{kl} + \genL_{ij} \big) \;.
\end{align}
We have omitted the time argument in \eqref{GeneralK} for clarity and used \eqref{DephasingGenerator}, \eqref{DephasingExponential}, \eqref{UnitaryGenerator}, and \eqref{UnitaryExponential} to obtain \eqref{GeneralL}.

\subsection{Application to experiment}

\begin{figure}[t]
\centerline{\includegraphics[width=0.25\textwidth]{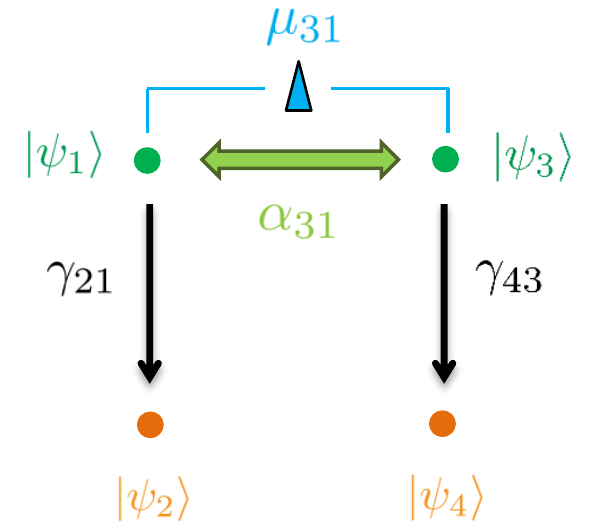}}
\caption{\label{GeneralGraph} A quantum walk involving all three processes described in Sec.~\ref{ReactOpFromQW}. Coherent evolution is represented by a green two-way arrow and parameterised by $\zeta_{31}$ (recall that the actual rate of oscillation is given by $\alpha_{31}$). This models the effect of the magnetic interactions such as the Zeeman or hyperfine. Any extra dephasing (i.e.~not caused by recombination) can be effectively described by the dephasing map. If applied to the experiment of Ref.~\cite{MLGH13} this would model the $g$-anisotropy for one of the radicals.}
\end{figure}

The experiment of Sec.~\ref{MaedaExpt} is an example where a model in the form of \eqref{GeneralK} and \eqref{GeneralL} would be directly applicable. Recall from Sec.~\ref{MaedaExpt} that the measured dephasing rate does not correspond to the dephasing caused by the recombination processes alone. Although the experiment minimises as many decoherent processes as possible not all such processes can be made negligible. We therefore include an additional source of dephasing in our model. This gives the resultant graph in Fig.~\ref{GeneralGraph}
whose time evolution can be described by the map
\begin{align}
	{\cal K}_{\rm QW}(dt) =  \mapV_{31}(dt) \, \mapU_{\;\!31}(dt) \, \mapM_{43}(dt) \, \mapM_{21}(dt)  \;.
\end{align}
Applying \eqref{GeneralK} and \eqref{GeneralL} gives the reaction operator:
\begin{align}
\label{GQW}
	{\cal G}_{\rm QW} \, \rho(t) 
	= \,& -i\big[ \hat{H}_{31} , \rho(t) \big]  \nn \\
	    & + k_{21} \bigg[ \hat{Q}_{21} \, \rho(t) \, \hat{Q}\dg_{21} - \frac{1}{2} \: \hat{Q}_{1} \, \rho(t) - \frac{1}{2} \: \rho(t) \: \hat{Q}_{1} \bigg]  \nn \\
	    & + k_{43} \bigg[ \hat{Q}_{43} \, \rho(t) \, \hat{Q}\dg_{43} - \frac{1}{2} \: \hat{Q}_{3} \, \rho(t) - \frac{1}{2} \: \rho(t) \: \hat{Q}_{3} \bigg]  \nn \\
      & + q_{31} \bigg[ \hat{Q}_{1} \, \rho(t) \, \hat{Q}_{1} - \frac{1}{2} \: \hat{Q}_{1} \, \rho(t) - \frac{1}{2} \: \rho(t) \: \hat{Q}_{1} \bigg]  \,.
\end{align}
The last line in \eqref{GQW} models any additional dephasing occurring on top of the recombination without contributing to population transfer (recall Sec.~\ref{CohEvoDeph}). Maeda and coworkers have suggested such a process \cite{MLGH13}. An order-of-magnitude estimate suggests the additional dephasing observed in their experiment can be accounted for by the anisotropy associated with the electron gyromagnetic ratio for one of the radicals (also referred to as $g$-anisotropy in electron paramagnetic resonance) \cite{MLGH13}. Their order-of-magnitude estimate for the rate of dephasing due to $g$-anisotropy could in principle be used in \eqref{GQW} for $q_{31}$. We will study of the effect of a nonzero $q_{31}$ in a sequel paper where a toy model is used to simulate radical-pair reactions in plant cryptochromes \cite{SCS07}. However, there we allow $q_{31}$ to be a free parameter so the dependece of the reaction kinetics on additional dephasing can be seen \cite{CGKPK14}.

We noted earlier in Sec.~\ref{AmpDampMapMainText} that the order of subscripts in the amplitude damping map is important as this determines the direction of the process. This is also the case for the dephasing map and is obviously true from \eqref{DephasingGenerator} as its generator is determined solely by the projector corresponding to the first index (counting from right to left). If we include products into our reaction operator then the form of \eqref{DephasingGenerator} will not only affect the singlet-triplet coherence of the radical pair, but also the coherences between the radical pair and products. The ordering of the indices then determines which coherences between the radical-pair and product are affected. Take the four-state graph in Fig.~\ref{GeneralGraph} for example, the contribution to the reaction operator given by $\mapV_{31}(dt)$ is
\begin{align}
\label{S31}
	{}& {\cal S}_{31}(q_{31}) \, \rho  \nn \\
	{}& = \left( \begin{array}{cccc} 
	  0                              &  -\half \, q_{31} \, \rho_{12}  &  -\half \, q_{31} \rho_{13}  &  -\half \, q_{31} \, \rho_{14}   \\ 
	  -\half \, q_{31} \, \rho_{21}  &  0                              &  0                           &  0                               \\
	  -\half \, q_{31} \rho_{31}     &  0                              &  0                           &  0                               \\
	  -\half \, q_{31} \, \rho_{41}  &  0                              &  0                           &  0
	\end{array}\right)  \;,
\end{align}
whereas the contribution from $\mapV_{13}(dt)$ is
\begin{align}
\label{S13}
	{}& {\cal S}_{13}(q_{13}) \, \rho  \nn \\
	{}& = \left( \begin{array}{cccc} 
	  0                              &  0                              &  -\half \, q_{13} \, \rho_{13}    &  0                               \\ 
	  0                              &  0                              &  -\half \, q_{13} \, \rho_{23}    &  0                               \\
	  -\half \, q_{13} \, \rho_{31}  &  -\half \, q_{13} \, \rho_{32}  &  0                                &  -\half \, q_{13} \, \rho_{34}   \\
	  0                              &  0                              &  -\half \, q_{13} \, \rho_{43}    &  0
	\end{array}\right) \;.
\end{align}
Note that we have explicitly written out the dependence on the dephasing rate in the generators to be clear. This therefore raises the question as to whether a dephasing map with a particular ordering of indices should be preferred over another, e.g.~$\mapV_{31}(dt)$ as opposed to $\mapV_{13}(dt)$ in \eqref{GQW}. It turns out that this asymmetry is usually not a problem for describing radical-pair reactions because the radical pair is usually created in either the singlet or the triplet state. More generally, the asymmetry in the dephasing map is irrelevant for any initially mixed state of the form
\begin{align}
\label{SymIniState}
	\rho(0) = \sum_{k=1}^N \, \wp_k \, \op{\psi_k}{\psi_k}  \;.
\end{align}
This is the case with the experiment of Sec.~\ref{MaedaExpt} where the radical pair begins in the singlet state, which corresponds to \eqref{SymIniState} with $N=4$, $\wp_1=1$, and $\wp_2=\wp_3=\wp_4=0$. This of course assumes that there are no other processes present which create coherences between the radical pair and products. We comment further on the asymmetry of the dephasing map and its use in deriving reaction operators below \footnote{There is an exception to this asymmetry when the graph has only two states. This can already be seen in \eqref{2StatePD} provided we set equal dephasing rates when reversing the order of indices. The two-state case is also equivalent to setting $\rho_{mn}=0$ except for $\rho_{13}$ in \eqref{S31} and \eqref{S13}. We can then make ${\cal S}_{31}(q_{31})={\cal S}_{13}(q_{13})$ by setting $q_{31}=q_{13}$.}.

\subsection{Variant models}

\subsubsection{Symmetric dephasing}

\begin{figure}[t]
\centerline{\includegraphics[width=0.18\textwidth]{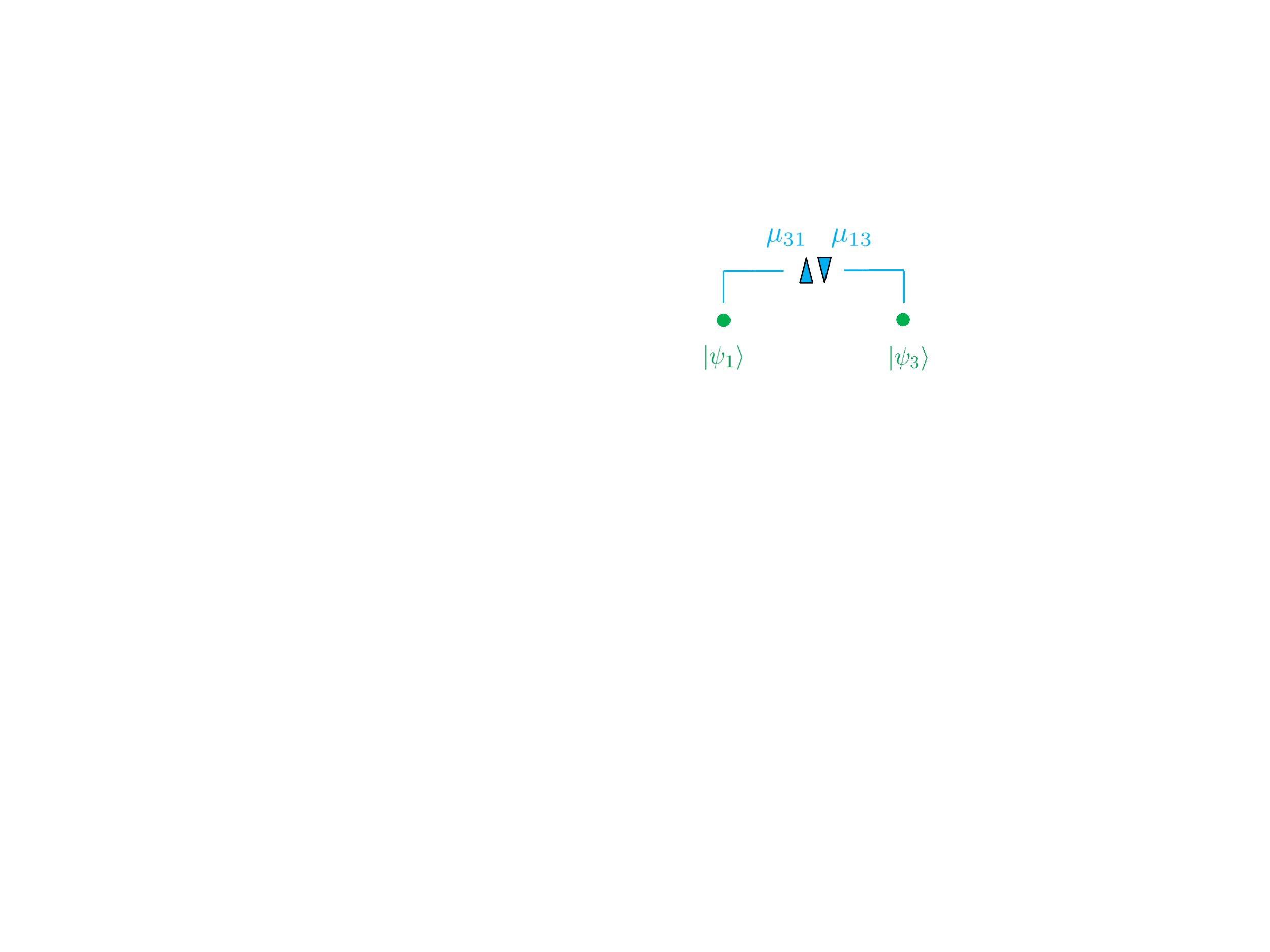}}
\caption{\label{KominisGraph} We represent two applications of the dephasing map by using multiple triangles with a single line. This gives an overall map that is symmetric in the decoherence rates when the rates for each direction are set equal. This is also the Kominis model with the dephasing rates replaced by the recombination rates.}
\end{figure}

If one insists that a symmetric dephasing model be used then this can be accomplished by two applications of the dephasing map. We represent such an operation by Fig.~\ref{KominisGraph}. Since we associate an upwards-pointing triangle with the map $\mapV_{31}(q_{31};dt)$ it makes sense to use a downwards-pointing triangle for $\mapV_{13}(q_{13};dt)$. We thus define the ``direction'' of dephasing by the direction in which the triangle points.
A symmetric dephasing map may then be defined by
\begin{align}
	\mapW_{31}(q;dt) \equiv \mapV_{31}(q_{31}:=q;dt) \, \mapV_{13}(q_{13}:=q;dt)  \;.
\end{align}
The generator corresponding to this is simply given by the sum of \eqref{S31} and \eqref{S13} with $q_{31} = q_{13} \equiv q$:
\begin{align}
\label{S13+S31}
	{\cal X}_{31}(q) \, \rho \equiv {}& \big[ {\cal S}_{13}(q) + {\cal S}_{31}(q) \big] \rho  \nn \\
	 = {}& \left( \begin{array}{cccc} 
	  0                          &  -\half \, q \, \rho_{12}  &  -q \, \rho_{13}             &  -\half \, q \, \rho_{14}        \\ 
	  -\half \, q \, \rho_{21}   &  0                         &  -\half \, q \, \rho_{23}    &  0                               \\
	  -q \, \rho_{31}            &  -\half \, q \, \rho_{32}  &  0                           &  -\half \, q \, \rho_{34}        \\
	  -\half \, q \, \rho_{41}   &  0                         &  -\half \, q \, \rho_{43}    &  0
	\end{array} \right) \;.
\end{align}

We can also provide a model of dephasing in which only the radical pair is referred to by using the partial trace defined earlier in Sec.~\ref{MainTextPartialTrace}. Since the partial trace of a sum is the sum of partial traces we can simply take the partial trace of \eqref{S13+S31} and add it to the generator for other processes. We can use exactly the same procedure as before to calculate the partial trace of \eqref{S13+S31} but it should be intuitive that the result is given by 
\begin{align}
\label{MatrixRhoRDeph}
	\dot{\rho}_{\rm R} 	
	{}& =	\left( \begin{array}{ccc} 
	  0  &  0                &  0                   \\ 
	  0  &  0                &  -q \, \rho_{31}     \\
	  0  &  -q \, \rho_{13}  &  0
	\end{array} \right) \;,	
\end{align}
where \eqref{MatrixRhoRDeph} is expressed in the $\{\ket{n_1,n_3}\}_{n_1,n_3}$ basis and the ordering of the matrix elements is the same as \eqref{MatrixRhoR}. Adding \eqref{MatrixRhoRDeph} to \eqref{MatrixRhoR} thus gives a model for the recombination process with added dephasing in the radical-pair basis $\{\ket{\rm S},\ket{\rm T},\ket{\rm N}\}$.

\subsubsection{Relation to Kominis's reaction operator}

This asymmetry in the dephasing map can be used to obtain Kominis's model operationally. Recall from Sec.~\ref{ReactOpModels} that his reaction operator is given by \eqref{LK} and is trace-preserving, with the radical-pair population evolved using a separate equation, given by \eqref{STPop}. This model can be seen as two applications of the dephasing map in opposite directions:
\begin{align}
	\rho(t+dt) = \mapV_{31}(q_{31}:=\ks;dt) \, \mapV_{13}(q_{13}:=\kt;dt) \, \rho(t)  \;.
\end{align}
Using \eqref{SitesST} this is equivalent to
\begin{align}
\label{KominisFromQW}
	 \frac{d\rho}{dt} = {}& {\cal S}_{31}(q_{31}:=\ks) \, {\cal S}_{13}(q_{13}:=\kt) \, \rho(t)  \nn \\[0.2cm]
                   = {}& \ks \bigg[ \Qs \, \rho(t) \, \Qs - \frac{1}{2} \: \Qs \, \rho(t) - \frac{1}{2} \: \rho(t) \: \Qs \bigg]     \nn \\
                       & + \kt \bigg[ \Qt \, \rho(t) \, \Qt - \frac{1}{2} \: \Qt \, \rho(t) - \frac{1}{2} \: \rho(t) \: \Qt \bigg]   \nn \\[0.2cm]
                   = {}&  \genL_{\rm K} \, \rho(t)  \;.
\end{align}
We can now understand why Kominis had to introduce an ad hoc method for describing population loss in the radical-pair reaction, namely that \eqref{KominisFromQW} only describes the loss of coherences. Of course we have not analysed the physics of Kominis's model. All we have done is to point out that whatever physics and assumptions go into Kominis's model, they amount to dephasing for the radical pair. This should be compared with the quantum-walk model used here in which the radical-pair dephasing is seen explicitly as a consequence of the decay out of $\ket{\psi_1}$ and $\ket{\psi_3}$ to products. If we want to describe population loss in the minimal basis $\{\ket{\rm S}, \ket{\rm T}\}$ then it makes more sense to use the conventional model given by $\genL_{\rm H}\rho$ (or equivalently $\bar{\genL}_{\rm QW}\rho$). We can of course also tack on $\genL_{\rm H}\rho$ to \eqref{KominisFromQW} to arrive at a model where there is population loss and additional dephasing in the minimal basis:
\begin{align}
	 \frac{d\rho}{dt} = \big( \genL_{\rm K} + \genL_{\rm H} \big) \, \rho(t)  \;.
\end{align}
The graph corresponding to Kominis's model is also given by Fig.~\ref{KominisGraph} when the dephasing rates are set equal to the recombination rates.


\section{Summary and Discussion}
\label{Conclusion}


We have shown that quantum walks provide the same level of treatment for coherent chemical kinetics as rate-equation models do for classical chemical kinetics by applying it to the radical-pair reaction of the avian compass. However the quantum walk considered here is not the same as those in the quantum-walk literature where an additional system degree of freedom is used as a coin \cite{APS12,APSS12}. The simplicity of our approach lies in the decomposition a multisite reaction into two-site processes. If multiple processes occur between two sites then we can also consider these processes separately. This gives a systematic method of deriving reaction operators. The breakdown into two-site processes also makes our approach less prone to modelling errors since two-site processes are much simpler to study---they are in fact well known in quantum information theory. As we saw in Sec.~\ref{ReactOpStdRPM} this allowed us to obtain a recombination dephasing rate for the standard radical-pair reaction which is consistent with experiments. This follows a similar line of thought as the Jones--Hore work except they attempted a derivation ``in one go''. Section~\ref{ReactOpStdRPM} can also be seen as a derivation of the Haberkorn reaction operator using a theory of coherent chemical kinetics as argued. We have also shown how the partial trace can be used to obtain a model where the products are ignored but still preserves the normalisation of the radical-pair state. This gives rise to a third state that has not been mentioned before in the radical-pair literature and can be seen as a half-way approach to the existing models (see Fig.~\ref{Summary}). We have also considered more general reactions in Sec.~\ref{MoreGeneralQWs} where a model corresponding to the experiment of Ref.~\cite{MLGH13} was introduced. We also considered variant forms of dephasing and discussed its relevance to the Kominis model.

Due to space we have not provided the details of how additional dephasing like the one in the model of \eqref{GQW} will affect a coherent reaction. The effect of such decoherent processes will ultimately depend on the actual values of the intermediate transition rates. In a sequel paper we further illustrate the use of quantum walks by simulating a toy model for a radical-pair reaction in plant cryptochromes where order-of-magnitude estimates for the intermediate transition rates are available \cite{SCS07}. However, instead of using a particular value of dephasing, we will vary the dephasing parameter over its full range so as to study the deviation of a maximally coherent reaction from one that is fully classical. For this purpose we use a well-known quantity called the hitting time from quantum walks as a measure of the quantumness of the reaction. This measures how quickly the reaction occurs as a function of the coherences in the reaction. Our sequel paper thus shows how the ``full'' machinery of quantum walks can be applied, not just the idea of composing maps presented in this paper. Such an application of coherent chemical kinetics can then be used to argue if a quantum description for the reaction being studied is necessary, or if simple classical rate equations will do.

Aside from the Jones--Hore work, the open-systems approach due to Tiersch and coworkers \cite{TSPB12} also deserves some comment. There is a sense in which our paper is similar to their's and a sense in which it is the complete opposite. Our work is similar to Ref.~\cite{TSPB12} in that Tiersch and colleagues also use maps to describe different sorts of evolution for the system (i.e.~the radical pair). Once the physics is seen clearly the system evolution is then described by a master equation, derived by converting the maps into differential form. However, whereas our goal is to establish the systematic use of maps for deriving reaction operators because of their operational nature, Tiersch and others use maps to suggest what underlying physics might give rise to known reaction operators. More precisely, Tiersch and colleagues treat the radical pair as an open system and ask what sort of system-environment interactions can give rise to sensible reaction operators. Their maps are thus different to our's in that their's act on the joint space of the system plus environment. In contrast, our maps act directly on the system alone. One can therefore say that our work and Ref.~\cite{TSPB12} are similar in methodology due to the simplicity of maps, but quite different in spirit.

\section{Acknowledgement}

We would like to thank Vlatko Vedral for useful discussions. TP acknowledges support from the Start-Up grant of the Nanyang Technological University and Ministry of Education grant number RG127/14. AC, PK, and DK acknowledge support from the National Research Foundation and Ministry of Education in Singapore. Work by LP was supported by the Polish Ministry of Science and Higher Education under the project number IP2012 051272.

\section*{Appendices}
\appendix


\section{Amplitude-damping Kraus map}
\label{AppAD}

\subsection{Operator-sum form}

For ease of reference we repeat the amplitude-damping map here:
\begin{align}
\label{AppAmpDampMap}
	\rho(t+\Dt) = {}& \mapM_{21} \, \rho(t)  \nn \\[0.2cm]
	            = {}& \hat{M}^{(1)}_{21}(\Dt) \, \rho(t) \big[ \hat{M}^{(1)}_{21}(\Dt) \big]\dg  \nn \\
	                & + \hat{M}^{(2)}_{21}(\Dt) \, \rho(t) \big[ \hat{M}^{(2)}_{21}(\Dt) \big]\dg  \;,
\end{align}
where the two Kraus operators are
\begin{gather}
\label{AppM1Dt}
	\hat{M}^{(1)}_{21}(\Dt) = \rt{\gamma_{21}(\Dt)} \, \op{\psi_2}{\psi_1}  \;,  \\
\label{AppM2Dt}
	\hat{M}^{(2)}_{21}(\Dt) = \op{\psi_2}{\psi_2} + \rt{1-\gamma_{21}(\Dt)} \, \op{\psi_1}{\psi_1}  \;.
\end{gather}
The forms of $\hat{M}^{(1)}_{21}(\Dt)$ and $\hat{M}^{(2)}_{21}(\Dt)$ can be understood by first noting that during $\Dt$ we either observe a transition or we do not (assuming of course that the system can be probed to give us this information). Considering first the case of observing a transition, the state is then
\begin{align}
\label{Rho1}
	\rho^{(1)}(t+\Dt) = \hat{M}^{(1)}_{21}(\Dt) \, \rho(t) \big[ \hat{M}^{(1)}_{21}(\Dt) \big]\dg / \wp_1(\Dt)  \;.
\end{align}
The probability of observing the transition is simply given by 
\begin{align}
	\wp_1(\Dt) = \gamma_{21}(\Dt) \rho_{11}(t)  \;,
\end{align}
where $\rho_{jj}(t)=\bra{\psi_j}\rho(t)\ket{\psi_j}$ is the probability of finding the system to be in $\ket{\psi_j}$ ($j=1,2$) at time $t$. It should be clear directly from the form of $\hat{M}^{(1)}_{21}(\Dt)$ that it effects the required conditional change as it is proportional to $\op{\psi_2}{\psi_1}$, whose effect is to take a system in state $\ket{\psi_1}$ to the state $\ket{\psi_2}$. The inclusion of $\rt{\gamma_{21}(\Dt)}$ then allows the correct probability of observing a transition to be found according to the formalism. Note the expression for $\wp_1(\Dt)$ is exactly what one would expect from classical probability theory. Considering now the case when we do not observe a transition, the state is then given by 
\begin{align}
\label{Rho2}
	\rho^{(2)}(t+\Dt) = \hat{M}^{(2)}_{21}(\Dt) \, \rho(t) \big[ \hat{M}^{(2)}_{21}(\Dt) \big]\dg / \wp_2(\Dt)  \;,
\end{align}
where
\begin{align}
	\wp_2(\Dt) = \rho_{22}(t) + \big[ 1 - \gamma_{21}(\Dt) \big] \rho_{11}(t)  \;,
\end{align}
is the probability of not observing a transition. These results can be understood by noting that two possible scenarios contribute to a no-transition observation: Either the system is already in state $\ket{\psi_2}$ at time $t$, which occurs with probability $\rho_{22}(t)$, or it is in state $\ket{\psi_1}$ but has not yet jumped to $\ket{\psi_2}$, which occurs with probability $[1-\gamma_{21}(\Dt)]\rho_{11}(t)$. The probability of not seeing a transition is therefore the sum of the probabilities for each of these scenarios. That $\hat{M}^{(2)}_{21}(\Dt)$ describes a combination of these two scenarios can also be seen from its form, which we can understand by a simple analogy to $\hat{M}^{(1)}_{21}(\Dt)$. Consider first the case where the system is in state $\ket{\psi_1}$ and remains in $\ket{\psi_1}$. Instead of taking $\ket{\psi_1}$ to $\ket{\psi_2}$ as in $\hat{M}^{(1)}(\Dt)$, we now take $\ket{\psi_1}$ to itself. This means that we simply replace the transition operator $\op{\psi_2}{\psi_1}$ in $\hat{M}^{(1)}(\Dt)$ by the projector $\op{\psi_1}{\psi_1}$. We would also have to replace the $\gamma_{21}(\Dt)$ under the square root in $\hat{M}^{(1)}(\Dt)$ by $1-\gamma_{21}(\Dt)$ since now we are concerned with the case where the system stays in $\ket{\psi_1}$. Doing so gives us $\hat{M}^{(2)}(\Dt)=\rt{1-\gamma_{21}(\Dt)}\op{\psi_1}{\psi_1}$, but we know this is not the complete description yet as we have not considered the contribution due to the system being in $\ket{\psi_2}$ and staying there. If the system is already in $\ket{\psi_2}$, the process should take $\ket{\psi_2}$ to itself because no other processes are present which can take the system out of $\ket{\psi_2}$. The probability that the system remains in $\ket{\psi_2}$ given that it was in $\ket{\psi_2}$ is thus simply 1. Therefore we simply add the projector $\op{\psi_2}{\psi_2}$ (with coefficient 1) to $\rt{1-\gamma_{21}(\Dt)}\op{\psi_1}{\psi_1}$ to arrive at the resultant form of $\hat{M}^{(2)}_{21}(\Dt)$. It is trivial to show that $\hat{M}^{(2)}_{21}(\Dt)$ produces the correct state by letting $\rho(t)$ be $\op{\psi_1}{\psi_1}$ and $\op{\psi_2}{\psi_2}$ in turn.

For the mathematically inclined reader, we note that \eqref{AppM2Dt} can be derived directly by using \eqref{AppM1Dt} and the constraint
\begin{align}
\label{ProbSum}
	\big[ \hat{M}^{(1)}_{21}(\Dt) \big]\dg \hat{M}^{(1)}_{21}(\Dt) + \big[ \hat{M}^{(2)}_{21}(\Dt) \big]\dg \hat{M}^{(2)}_{21}(\Dt) = \hat{1}  \;.
\end{align}
By resolving the identity on the right-hand side in the site basis, \eqref{ProbSum} gives
\begin{align}
\label{M2dagM2}
	\big[ \hat{M}^{(2)}_{21}(\Dt) \big]\dg \hat{M}^{(2)}_{21}(\Dt) = {}& \big[ 1-\gamma_{21}(\Dt) \big] \, \op{\psi_1}{\psi_1}  \nn \\
	                                                                   & + \op{\psi_2}{\psi_2}    \;.
\end{align}
Note that $\hat{M}^{(2)}_{21}(\Dt)$ is simply the operator square root of this equation. Since \eqref{M2dagM2} is diagonal, we arrive at \eqref{AppM2Dt} on taking the sqaure root of the coefficients of $\op{\psi_1}{\psi_1}$ and $\op{\psi_2}{\psi_2}$. This derivation of \eqref{AppM2Dt} is simple but does contain the insight provided above.

Now that we have the necessary Kraus operators the evolution of the system follows directly by forming the sum \eqref{AppAmpDampMap}. Notice from the above that $\hat{M}^{(1)}_{21}(\Dt)$ and $\hat{M}^{(2)}_{21}(\Dt)$ are essentially time-evolution operators but conditioned on our knowledge of whether the system underwent a jump or not. Equations \eqref{Rho1} and \eqref{Rho2} are in fact quantum analogues of the classical Bayes rule \cite{SBC01}. Conditioning requires that we monitor the system for the entire duration of $\Dt$. What the operator-sum representation of $\rho(t+\Dt)$ describes is how the state should evolve without us having to monitor the system continuously, or in the language of probability theory, it describes the unconditioned state. This can be understood as follows: If one does not monitor the system then all we can say is that with probability $\wp_1(\Dt)$ the system will be in the state $\rho^{(1)}(t+\Dt)$, and with probability $\wp_2(\Dt)$ the system will be in the state $\rho^{(2)}(t+\Dt)$. From probability theory, we would say that the state in the absence of such monitoring at time $t+\Dt$ (i.e.~the unconditioned state) is therefore a weighted sum of the conditioned states $\rho^{(1)}(t+\Dt)$ and $\rho^{(2)}(t+\Dt)$,
\begin{align}
\label{AvgState}
	\rho(t+\Dt) = \wp_1(\Dt) \, \rho^{(1)}(t+\Dt) + \wp_2(\Dt) \, \rho^{(2)}(t+\Dt)  \;.
\end{align}
Substituting in \eqref{Rho1} and \eqref{Rho2} we obtain exactly the result of Kraus. In practice one often has an ensemble of particles and all we know is the fraction of particles that underwent a state transition during $\Dt$. In this case $\wp_1(\Dt)$ is simply the fraction of particles that jumped and $\wp_2(\Dt)$ the fraction that did not. Note also the difference between $\wp_1(\Dt)$ and $\gamma_{21}(\Dt)$---The former is the probability of observing a transition from $t$ to $t+\Dt$ without assuming that we know which state the system is in at time $t$, whereas the latter does, $\gamma_{21}(\Dt)$ is the probability of jumping conditioned on the system being in state $\ket{\psi_1}$ at time $t$.

\subsection{Differential form}

We have now described a simple one-way population transfer completely as a probabilistic process. Instead of expressing the system evolution as a sum over conditioned states we can express it in the form of a differential equation. Such an equation can be derived by considering the evolution of $\rho(t)$ over an infinitesimal interval dt. In this case it is more appropriate to refer to the rate at which the system jumps from $\ket{\psi_1}$ to $\ket{\psi_2}$ over some interval $\Dt$ rather than the probability $\gamma_{21}(\Dt)$. If we denote the fraction of particles that jump from $\ket{\psi_1}$ to $\ket{\psi_2}$ per second by $k_{21}$, the probability $\gamma_{21}(\Dt)$ is then related to $k_{21}$ by $\gamma_{21}(\Dt)=k_{21}\Dt$. This means that the probability of a jump in an infinitesimally small time interval is also an infinitesimal. The evolution of the system state is now given by
\begin{align}
\label{AmpDampRho(t+dt)}
	\rho(t+dt) = {}& \hat{M}^{(1)}_{21}(dt) \, \rho(t) \big[ \hat{M}^{(1)}_{21}(dt) \big]\dg  \nn \\
	                & + \hat{M}^{(2)}_{21}(dt) \, \rho(t) \big[ \hat{M}^{(2)}_{21}(dt) \big]\dg  \;,
\end{align}
Using the binomial expansion we have
\begin{align}
\label{Binomial}
	\rt{1-\gamma_{21}(dt)} = 1 - \frac{k_{21}}{2} \, dt  \;.
\end{align}
We can then write the Kraus operators for infinitesimal evolution using $\hat{Q}_{21}=\op{\psi_2}{\psi_1}$ and $\hat{Q}_1=\op{\psi_1}{\psi_1}$ as
\begin{gather}
\label{AmpDampOp1dt}
	\hat{M}^{(1)}_{21}(dt) = \rt{k_{21} \,dt} \, \hat{Q}_{21}  \;,  \\
\label{AmpDampOp2dt}
	\hat{M}^{(2)}_{21}(dt) = \hat{1} - \frac{k_{21}}{2} \: \hat{Q}_{1} \, dt  \;.
\end{gather}
Substituting \eqref{AmpDampOp1dt} and \eqref{AmpDampOp2dt} into \eqref{AmpDampRho(t+dt)} and neglecting terms on the order of $dt^2$ we arrive at 
\begin{align}
\label{dp21/dt}
	\frac{d\rho}{dt} = k_{21} \bigg[ \hat{Q}_{21} \, \rho(t) \, \hat{Q}\dg_{21} - \frac{1}{2} \: \hat{Q}_{1} \, \rho(t) - \frac{1}{2} \: \rho(t) \, \hat{Q}_{1} \bigg]  \;.
\end{align}
This is the master equation corresponding to the amplitude damping map and can be put in the Lindblad form if one wishes by using the property $\hat{Q}_1=\hat{Q}\dg_{21} \, \hat{Q}_{21}$. We have derived this equation rather simply by applying the Kraus formalism, hence it can be regarded as merely a restatement of the operator-sum representation of $\rho(t)$ in differential form. It is only a matter of preference whether one wants to use a map or a differential equation to simulate the system dynamics but the Kraus formalism provides a simple way to understand the essential physics of the process by using only basic probability ideas.


\section{Partial trace over chemical products}
\label{PartialTrace}

Here we show how to obtain an equation of motion for the radical pair where the products are ignored by using the partial trace operation on \eqref{LQW}. We have already argued in Sec.~\ref{MainTextPartialTrace} that this is given by 
\begin{align}
\label{AppRhoR}
	\dot{\rho}_{\rm R}(t) = {}& \bra{n_2=0,n_4=0} \, \dot{\rho}(t) \, \ket{n_2=0,n_4=0} \nn \\
                            & + \bra{n_2=0,n_4=1} \, \dot{\rho}(t) \, \ket{n_2=0,n_4=1}  \nn \\
                            & + \bra{n_2=1,n_4=0} \, \dot{\rho}(t) \, \ket{n_2=1,n_4=0}  \;.
\end{align}
where
\begin{align}
\label{AppLQW}
	\dot{\rho}(t) = {}& k_{21} \Bigg[ \hat{Q}_{21} \, \rho(t) \, \hat{Q}\dg_{21} 
	                    - \frac{1}{2} \: \hat{Q}_{1} \, \rho(t) - \frac{1}{2} \: \rho(t) \: \hat{Q}_{1} \Bigg]  \nn \\
	                  & + k_{43} \Bigg[ \hat{Q}_{43} \, \rho(t) \, \hat{Q}\dg_{43}  
	                    - \frac{1}{2} \: \hat{Q}_{3} \, \rho(t) - \frac{1}{2} \: \rho(t) \: \hat{Q}_{3} \Bigg] \;,
\end{align}
and we have written the site basis using $\ket{n_1,n_2,n_3,n_4}$ as [recall \eqref{Psi1}--\eqref{Psi4} and \eqref{Tensor}] 
\begin{align}
	\ket{\psi_1} = \ket{1,0,0,0} \;,  \quad  	\ket{\psi_3} = \ket{0,0,1,0}  \;, \\
	\ket{\psi_2} = \ket{0,1,0,0} \;,  \quad  	\ket{\psi_4} = \ket{0,0,0,1}  \;.
\end{align}

The sum \eqref{AppRhoR} is most easily calculated by first rewriting \eqref{AppLQW} as
\begin{align}
\label{RhoRewritten}
	\dot{\rho} = {}& k_{21} \Bigg( \rho_{11} \, \op{\psi_2}{\psi_2} - \frac{1}{2} \sum_{m=1}^4 \, \rho_{1m} \, \op{\psi_1}{\psi_m}  \nn \\
	               & - \frac{1}{2} \sum_{m=1}^4 \rho_{m1} \, \op{\psi_m}{\psi_1} \Bigg) + k_{43} \Bigg( \rho_{33} \, \op{\psi_4}{\psi_4}  \nn \\
	               & - \frac{1}{2} \sum_{m=1}^4 \, \rho_{3m} \, \op{\psi_3}{\psi_m} - \frac{1}{2} \sum_{m=1}^4 \rho_{m3} \, \op{\psi_m}{\psi_3} \Bigg) \;.
\end{align}
The following identities will therefore prove useful
\begin{align}
\label{TraceId1}
	{}& \ip{n_2=0,n_4=0\,}{\psi_m}  \nn \\
	{}& = (1-\delta_{m2}) (1-\delta_{m4}) \, \ket{n_1=\delta_{m1},n_3=\delta_{m3}}  \;,  \\[0.2cm]
\label{TraceId2}
	{}& \ip{n_2=0,n_4=1\,}{\psi_m}  \nn \\
	{}& =  (1-\delta_{m2}) \, \delta_{m4} \, \ket{n_1=\delta_{m1},n_3=\delta_{m3}}  \;,  \\[0.2cm]
\label{TraceId3}
	{}& \ip{n_2=1,n_4=0\,}{\psi_m}  \nn \\
	{}& =  \delta_{m2} \, (1-\delta_{m4}) \, \ket{n_1=\delta_{m1},n_3=\delta_{m3}}  \;. 
\end{align}
We will also use the summation convention where a repeated index is summed over. Using \eqref{RhoRewritten} the first term in \eqref{AppRhoR} can thus be calculated as follows.
\begin{align}
\label{TermA}
	{}& \bra{n_2=0,n_4=0} \, \dot{\rho}(t) \, \ket{n_2=0,n_4=0}  \nn \\
	{}& = - \frac{k_{21}}{2} \; \Big( \rho_{1m} \, \ip{n_2=0,n_4=0\,}{\psi_1} \ip{\psi_m}{n_2=0,n_4=0}  \nn \\
	  & \quad + \rho_{m1} \, \ip{n_2=0,n_4=0\,}{\psi_m} \ip{\psi_1}{n_2=1,n_4=0} \Big) \nn \\
	  & \quad - \frac{k_{43}}{2} \, \Big(  \rho_{3m} \, \ip{n_2=0,n_4=0\,}{\psi_3} \ip{\psi_m}{n_2=0,n_4=0}  \nn \\
	  & \quad + \rho_{m3} \, \ip{n_2=0,n_4=0\,}{\psi_m} \ip{\psi_3}{n_2=0,n_4=0} \Big) \;,
\end{align}
where we have noted on using \eqref{TraceId1} that
\begin{align}
	\ip{n_2=0,n_4=0\,}{\psi_2} = \ip{n_2=0,n_4=0\,}{\psi_4} = 0  \;.
\end{align}
while the first term in \eqref{TermA} is 
\begin{align}
	{}& \rho_{1m} \, \ip{n_2=0,n_4=0\,}{\psi_1} \ip{\psi_m}{n_2=0,n_4=0}  \nn \\
	{}& = \rho_{1m} \, (1 - \delta_{m2}) \, (1 - \delta_{m4}) \nn \\
	& \quad \times \op{n_1=1,n_3=0}{n_1=\delta_{m1},n_3=\delta_{m3}}  \nn \\[0.2cm]
	{}& = \rho_{1m} \, ( 1 - \delta_{m4} - \delta_{m2} - \cancel{\delta_{m2} \, \delta_{m4}} )  \nn \\
	  & \quad \times \op{n_1=1,n_3=0}{n_1=\delta_{m1},n_3=\delta_{m3}}  \nn \\[0.2cm]
	{}& = \ket{n_1=1,n_3=0} \Big( \rho_{1m} \, \bra{n_1=\delta_{m1},n_3=\delta_{m3}}  \nn \\
	  & \quad - \delta_{m4} \, \rho_{1m} \, \bra{n_1=\delta_{m1},n_3=\delta_{m3}}  \nn \\
	  & \quad - \delta_{m2} \, \rho_{1m} \, \bra{n_1=\delta_{m1},n_3=\delta_{m3}} \, \Big)  \nn \\[0.2cm]
	{}& = \ket{n_1=1,n_3=0} \Big( \rho_{11} \, \bra{n_1=1,n_3=0} \nn \\
	  & \quad + \cancel{\rho_{12} \, \bra{n_1=0,n_3=0}} + \rho_{13} \, \bra{n_1=0,n_3=1} \nn \\
	  & \quad + \cancel{\rho_{14} \, \bra{n_1=0,n_3=0}} - \cancel{\rho_{14} \, \bra{n_1=0,n_3=0}} \nn \\
	  & \quad - \cancel{\rho_{12} \, \bra{n_1=0,n_3=0}} \, \Big)  \nn \\[0.2cm]
\label{A2}
	{}& = \rho_{11} \, \op{n_1=1,n_3=0}{n_1=1,n_3=0}  \nn \\
	  & \quad - \rho_{13} \, \op{n_1=1,n_3=0}{n_1=0,n_3=1}  \;.
\end{align}
This also gives us the second term in \eqref{TermA} since it is just the Hermitian conjugate of \eqref{A2},
\begin{align}
\label{A3}
	{}& \rho_{m1} \, \ip{n_2=0,n_4=0\,}{\psi_m} \ip{\psi_1}{n_2=0,n_4=0}  \nn \\
	{}& = \Big( \rho_{1m} \, \ip{n_2=0,n_4=0\,}{\psi_1} \ip{\psi_m}{n_2=0,n_4=0} \Big)\dg  \nn \\[0.2cm]
	{}& = \rho_{11} \, \op{n_1=1,n_3=0}{n_1=1,n_3=0}  \nn \\
	  & \quad - \rho_{31} \, \op{n_1=0,n_3=1}{n_1=1,n_3=0}	\;.
\end{align}

The third term in \eqref{TermA} can be calculated in exactly the same manner with $\rho_{3m}$ replacing $\rho_{1m}$, and $\ket{\psi_3}$ replacing $\ket{\psi_1}$. By inspection of the above working we see that this amounts to making the following replacements in \eqref{A2} (bras remain unchanged):
\begin{gather}
	\rho_{11} \longrightarrow \rho_{31} \;, \quad \rho_{13} \longrightarrow \rho_{33}  \;,   \\
	\ket{n_1=1,n_3=0} \longrightarrow \ket{n_1=0,n_3=1} \;. 
\end{gather}
We thus obtain
\begin{align}
\label{A4}
	{}& \rho_{3m} \, \ip{n_2=0,n_4=0\,}{\psi_3} \ip{\psi_m}{n_2=0,n_4=0}  \nn \\
	{}& = \rho_{31} \, \op{n_1=0,n_3=1}{n_1=1,n_3=0}  \nn \\
	  & \quad - \rho_{33} \, \op{n_1=0,n_3=1}{n_1=0,n_3=1}  \;,
\end{align}
and taking the Hermitian conjugate,
\begin{align}
\label{A5}
	{}& \rho_{m3} \, \ip{n_2=0,n_4=0\,}{\psi_m} \ip{\psi_3}{n_2=0,n_4=0}  \nn \\
	{}& = \Big( \rho_{3m} \, \ip{n_2=0,n_4=0\,}{\psi_3} \ip{\psi_m}{n_2=0,n_4=0} \Big)\dg  \nn \\[0.2cm]
	{}& = \rho_{13} \, \op{n_1=1,n_3=0}{n_1=0,n_3=1}  \nn \\
	  & \quad - \rho_{33} \, \op{n_1=0,n_3=1}{n_1=0,n_3=1}	\;.
\end{align}
Substituting \eqref{A2}, \eqref{A3}, \eqref{A4}, and \eqref{A5} into \eqref{TermA} and collecting like terms we arrive at
\begin{align}
\label{R1}
	{}& \bra{n_2=0,n_4=0} \, \dot{\rho}(t) \, \ket{n_2=0,n_4=0}  \nn \\
	{}& = - \, k_{21} \, \rho_{11} \, \op{n_1=1,n_3=0}{n_1=1,n_3=0}  \nn \\
	  & \quad - \frac{1}{2} \; \big( k_{21} + k_{43} \big) \rho_{13} \, \op{n_1=1,n_3=0}{n_1=0,n_3=1}  \nn \\
	  & \quad - \frac{1}{2} \; \big( k_{21} + k_{43} \big) \rho_{31} \, \op{n_1=0,n_3=1}{n_1=1,n_3=0}  \nn \\
	  & \quad - k_{43} \, \rho_{33} \, \op{n_1=0,n_3=1}{n_1=0,n_3=1}  \;.
\end{align}

The remaining two terms in \eqref{RhoR} are much easier to calculate with the help of \eqref{TraceId2} and \eqref{TraceId3}. The second term in \eqref{AppRhoR} is given by
\begin{align}
\label{R2}
	{}& \bra{n_2=0,n_4=1} \, \dot{\rho}(t) \, \ket{n_2=0,n_4=1}  \nn \\
	{}& = k_{43} \, \rho_{33} \, \op{n_1=0,n_3=0}{n_1=0,n_3=0}  \;,
\end{align}
where we have noted from \eqref{TraceId2} that
\begin{align}
	\ip{n_2=0,n_4=1\,}{\psi_m} = {}& \ket{n_1=0,n_3=0} \;\, \text{for $m = 4$}  \;, \nn \\
	                           = {}& 0 \;\, \text{for $m = 1,2,3$} \;.
\end{align}
and 
\begin{align}
	\ip{n_2=0,n_4=1\,}{\psi_4} = \ket{n_1=0,n_3=0} \;.
\end{align}
Similarly, the third term in \eqref{AppRhoR} is
\begin{align}
\label{R3}
	{}& \bra{n_2=1,n_4=0} \, \dot{\rho}(t) \, \ket{n_2=1,n_4=0}  \nn \\
	{}& = k_{21} \, \rho_{11} \, \op{n_1=0,n_3=0}{n_1=0,n_3=0}  \;,
\end{align}
where we have noted from \eqref{TraceId3} that
\begin{align}
	\ip{n_2=1,n_4=0\,}{\psi_m} = {}& \ket{n_1=0,n_3=0} \;\, \text{for $m=2$} \;,  \nn \\
	                           = {}& 0 \;\, \text{for $m = 1,3,4$} \;.
\end{align}
Collecting \eqref{R1}, \eqref{R2}, and \eqref{R3}, we finally have the state $\rho_{\rm R}$ which has the products traced out.  The final form of \eqref{AppRhoR} is then 
\begin{align}
\label{FinalRhoR}
	\dot{\rho}_{\rm R} = {}& \big( k_{21} \, \rho_{11} + k_{43} \, \rho_{33} \big) \op{0,0}{0,0} - k_{43} \, \rho_{33} \, \op{0,1}{0,1}  \nn \\
	                       & - k_{21} \, \rho_{11} \, \op{1,0}{1,0} - \frac{1}{2} \; \big( k_{21} + k_{43} \big) \, \rho_{13} \, \op{1,0}{0,1}  \nn \\ 
	                       & - \frac{1}{2} \; \big( k_{21} + k_{43} \big) \, \rho_{31} \, \op{0,1}{1,0}  \;.
\end{align}
where we have used the fact that $\rho_{\rm R}$ must be spanned by $\{\ket{n_1,n_3}\}_{n_1,n_3}$ to omit writing out $n_1$ and $n_3$ explicitly in \eqref{FinalRhoR}.

\end{document}